\documentclass[a4paper,12pt]{article}
\usepackage[utf8]{inputenc}
\usepackage[italian, english]{babel}
\usepackage[left=2cm,right=2cm,top=2cm,bottom=2cm]{geometry}

\usepackage{amsmath} %necessario per matrici
\usepackage{graphicx} %necessario per immagini
\usepackage{float}
\usepackage{amsthm} %necessario per teoremi/lemmi ecc
\usepackage{amsfonts} %necessario per mathbb e alcuni simboli tra cui il quadrato
\usepackage{amssymb} %necessario per fare \hslash
\usepackage{xcolor} %necessario per i colori
\usepackage{listings} %necessario per scrivere i codici fai attenzione che nel codice non compaiano termini accentati o ti da errore
\definecolor{backcolour}{rgb}{0.95,0.95,0.92}
\usepackage{verbatim}
\usepackage{siunitx}
\usepackage{youngtab}
\usepackage{ytableau}
\usepackage{subfig}
\usepackage{braket}
\usepackage{physics}
\usepackage{slashed}
\usepackage{booktabs}
 \usepackage{mathrsfs}
 \usepackage{bbm}
\usepackage{tikz}
\usepackage{dynkin-diagrams} 
\usepackage[T1]{fontenc} % for \symbol{92} 
\usepackage{yfonts}

\usepackage[colorlinks=true, linkcolor=blue, citecolor=blue]{hyperref} %necessario per i collegamenti IMPORTA PER ULTIMO

\numberwithin{equation}{section}

\usepackage{bbm}
\usepackage[utf8]{inputenc}
\usepackage{graphicx}
\usepackage{hyphenat}
\usepackage{wrapfig}
\usepackage{empheq}
\usepackage{textcomp}
\usepackage{rotating}
\usepackage{wrapfig}
\usepackage{pdfpages}
\usepackage{verbatim}
\usepackage{setspace}
\usepackage{array}
\usepackage{caption}
\usepackage[pdf]{pstricks}
\usepackage{upgreek}
\usepackage{cmll}
\usepackage{latexsym}
\usepackage{braket}
\usepackage{cite}
\usepackage{epsfig}

\usepackage{tikz-cd}

\newcommand{\beq}{\begin{eqnarray}}
\newcommand{\eeq}{\end{eqnarray}}
\newcommand{\bea}{\begin{eqnarray}}
\newcommand{\eea}{\end{eqnarray}}
\newcommand{\be}{\begin{equation}}
\newcommand{\ee}{\end{equation}}

\newcommand{\im}{\mathrm{i}}

\newcommand{\rme}{\mathrm{e}}

\def\brc{\langle }
\def\ckt{\rangle}
\def\const{{\rm const}}
   
\def\de{\partial}
\numberwithin{equation}{section}

\begin{document}

%\pagestyle{myheadings}
%\markright{Electroweak Theory }

%\parindent 0mm
\parskip 4pt

% Start with the article header information
% Include all author details, including postal and e-mail addresses

\title{Infrared spectra of some\\ strongly--coupled chiral gauge theories}

\author{
Stefano Bolognesi$^{(1,2)}$, 
  Kenichi Konishi$^{(1,2)}$,   
   Matteo Orso$^{(1,2)}$    
  \\[13pt]
{\em \footnotesize
$^{(1)}$Department of Physics ``E. Fermi'', University of Pisa,}\\[-5pt]
{\em \footnotesize
Largo Pontecorvo, 3, Ed. C, 56127 Pisa, Italy}\\[2pt]
{\em \footnotesize
$^{(2)}$INFN, Sezione di Pisa,    
Largo Pontecorvo, 3, Ed. C, 56127 Pisa, Italy}\\[2pt]
\\[1pt] 
{ \footnotesize  stefano.bolognesi@unipi.it,  kenichi.konishi@unipi.it, matteo.orso@phd.unipi.it }  
} 
\date{}

\maketitle

\begin{abstract}
Several simple asymptotically-free chiral gauge theories are studied.  The only ``free parameters'' of our models are the choice of the gauge group
and the matter Weyl fermion representations, and the relative magnitudes of the renormalization-group-invariant scales $\Lambda_i$ associated with  each gauge group. 
None of our models has nontrivial nonabelian global symmetries (``family''--like fermion representations).
 We rely on some recent theoretical developments
on the dynamics of strongly--coupled chiral gauge theories, based on the generalized symmetries and associated new types of anomaly-matching consideration, but also on the solid knowledge on vectorlike gauge theories such as QCD and supersymmetric Yang-Mills theories.  The  structures of the infrared effective theories, the RG flows,  and the light spectra found in these models are surprisingly rich and intriguing.

\end{abstract}

\maketitle

\newpage

\tableofcontents

%\newpage

\section{Introduction}

   Our understanding of the dynamics of strongly--coupled gauge theories in four dimensions is still quite preliminary. 
Exceptions are the vectorlike theories such as QCD and ${\cal N}=2$  supersymmetric gauge theories, where many detailed results are known either  
by numerical (the lattice) or exact, analytical methods.  Unfortunately, in the case of chiral gauge theories, which could possibly be relevant to model building beyond the Standard Model,  our quantitative grip of their properties is still quite poor. 
One feels that it is necessary to make our theoretical tools more powerful,  before we will be able to take concrete, realistic steps in building the
correct theory of the fundamental interactions, which reduces at low energies to the Standard $SU(3)\times SU(2)_L \times U(1)_Y$ model with the known quarks and leptons and their properties,  and accounting for the Higgs sector in a totally natural fashion. 
The aim of the present work is to try to make a few modest steps beyond the earlier
\cite{Raby,Dimopoulos:1980hn,BY,Eichten:1985ft,Goity:1985tf,Eichten:1985fs,ACSS,ADS,Poppitz,AS,Appelquist:2013oni,ShiShr1,ShiShr2,Ryttov:2017gtm,th,Csaki:1997aw}
and more recent series of investigations  \cite{BKS,BK,Yamaguchi,BKL1,BKL2,BKL4,BKL5,BKLReview,BKLDA,BKLproc,BKLZ2,corfu,AnberHong,AnberChan,BKLO1,BKLO2}   on the  dynamics of chiral gauge theories.

The models considered here contain  a set of nonabelian gauge fields and fermions in some simple representations. The gauge interactions are all asymptotically-free (AF). The only freedom we have, apart from the choice of the gauge groups and of the matter fermions,  
is the relative ordering of the renormalization group (RG) -- invariant mass scales of each gauge group, $\{\Lambda_i\}$,  $i=1,2,\ldots$.

Our models do not possess any
 global symmetry groups, except perhaps some $U(1)$  groups,  which are possible Adler--Bell--Jackiw (ABJ) anomaly--free combinations of the classical fermion--number groups, associated with each type of fermion. This occurs in the first model considered below.  A priori, these $U(1)$ symmetries could have 't Hooft anomalies, but actually this turns out not to be the case.   The conventional 't Hooft anomaly-matching consideration does not play a central role in this work, in finding out the dynamics of the systems as they flow towards the infrared (IR).    
 
We wish to find out, given the model,  which type of low-energy effective theories can result dynamically, and which kind of light spectrum and small mass parameters 
emerge, depending  on  the relative values of  $\{\Lambda_i\}$.  Whichever low-energy effective theory and particle spectra we find in each case  will be  totally natural, in the sense that  all radiative corrections and renormalizations eventually (in principle) lead back in the ultraviolet (UV), to the original, simple asymptotically-free (hence, free) theory. 

 This paper is organized as follows. 
In Sec. \ref{sec:2} we consider an $SU(N)-SU(N+4)$ model which is a sort of combination of the so--called Bars--Yankielowicz  and Georgi--Glashow  models. In Sec. \ref{quiver},   quiver $SU(N)^n$ models are studied.  Sec. \ref{sec:4}  examines  an $SU(N)-Sp(6)-Sp(6)$ model, which is a generalization of the Pati-Salam model.     In Sec. \ref{AFIF} we analyze  an $SU(10)$ gauge theory with a single Weyl fermion in the  fifth--rank antisymmetric (self--adjoint) representation.  Sec. \ref{summary} summarizes what are found in these models.

\section{The $SU(N)-SU(N+4)$  model   \label{sec:2} }

The first model we consider here is an $SU(N)\times SU(N+4) $ gauge theory  with left-handed fermions shown in Table~\ref{Simplest}.
\begin{table}[h!t]
  \centering 
  \begin{tabular}{|c|c |c|c| }
\hline  
 ${}_{\rm fields} \backslash {}^{\rm groups}$ &  $SU(N) $    &  $ SU(N+4)$   &    $U(1)$   \\ 
 \hline 
   &&&     \\[-2ex]
         $\psi$     &     
     $ \Yvcentermath1 { \yng(2)} $  &    $  {\underline {1}} $    &    $N+4$    \\[1.6ex]
  $ \eta $      &   $ \Yvcentermath1  {\bar  {\yng(1)}}$     &  $ \Yvcentermath1  {\bar  {\yng(1)}}$  &   $- (N+2)$     \\[1.6ex]
  $ \chi $      &   $  {\underline {1}} $      &   $ \Yvcentermath1 { \yng(1,1)} $     &   $N$    \\[1.6ex]
\hline
\end{tabular}  
  \caption{\footnotesize    The gauge groups and matter (Weyl) fermions.  $U(1)$ is the unique, anomaly-free combination of the global fermion number symmetries.  }
   \label{Simplest}
\end{table}
The gauge interactions become stronger towards the infrared.  
The infrared spectra of the model will be different, depending on which of the   $SU(N+4)$ or $SU(N)$ gauge interactions is strongest first along the RG flow.

When  $g_{SU(N+4)} = 0$,  the model is the well-known ``$\psi\eta$''  (Bars--Yankielowicz) model investigated in \cite{BY,ACSS,ADS,BKS,BKL2, BKL4, BKL5, BKLReview, BKLproc, BKLZ2, corfu, BKLO1}, with free $\chi$ fermions
that act as spectator fermions to make  the global $SU(N+4)$ anomaly-free.  If instead  $g_{SU(N)} = 0$,  the model is the ``$\chi\eta$''
(Georgi--Glashow) model  studied in \cite{ADS,BKS,BKL4,BKL5,BKLReview,BKLproc,BKLZ2,corfu, BKLO1}, with  extra free spectator fermions  $\psi$.

 The global  $U(1)$ symmetry is free of ABJ  anomalies  with respect to $SU(N+4)$ or  to   
$SU(N)$.  Also, the  $[U(1)]^3$ as well as  $[U(1)]-[{\rm gravity}]^2$ anomalies turn out to cancel, hence do not introduce any significant anomaly-matching constraint
on the possible infrared dynamics.

\subsection{$SU(N)$ interactions getting strong first \label{SUNfirst} } 

Let us first consider the case where the $SU(N)$ interactions become strong at some scale $\Lambda_1$.  We assume that at that mass scale  $SU(N+4)$
interactions are weak and can be treated perturbatively.  In the limit  $g_{SU(N+4)} \to 0$, the model reduces to   the Bars-Yankielowicz  model, with a free extra fermions $\chi$.  
In other words,  
\be   g_1 (\equiv g_{SU(N)})    \gg       g_2  (\equiv  g_{SU(N+4)})  \;,\label{dueto} 
\ee
in the UV.
Recent observations based on a ${\mathbb{Z}}_{2}$  anomaly \cite{BKL2,BKL4,BKLZ2,corfu}  (see more below)  suggest that  when   the  $\psi\eta$ system  becomes strongly--coupled at the mass scale $\sim \Lambda_1$,  it goes into dynamical Higgs phase with ``color--flavor locked'' bifermion condensate formation  \cite{Raby,BY,ACSS,ADS},
\be   \brc \psi^{ij} \eta_{j\,a}  \ckt=  \delta^i_a  \, (\Lambda_1)^2  \;, \qquad   i,j=1,2,\ldots N, \quad a=1,2,\ldots, N+4\;.    \label{VEV1}  
\ee 
The fermions  in the UV and in the IR  are shown  in Table~\ref{SimplestBis},   taken from \cite{BKL2}.
\begin{table}[h!t]
{
  \centering 
  \begin{tabular}{|c|c|c |c|c|c|  }
\hline
$ \phantom{{{   {  {\yng(1)}}}}}\!  \! \! \! \! \!\!\!$   & ${}_{\rm fields} \backslash {}^{\rm groups}$   &  $SU(N)$    &  $ SU(4)$     &   $  U(1)^{\prime}   $   &  $({\mathbb{Z}}_{2})_F$    \\
 \hline
   \phantom{\huge i}$ \! \!\!\!\!$  {\rm UV}&  $\psi$   &   $\Yvcentermath1 { \yng(2)} $  &    $  \frac{N(N+1)}{2} \cdot   {\underline {1}}$    & $   N+4  $   & $1$  \\[1.8ex]
 & $ \eta_{a_1}$      &   $ \Yvcentermath1 {\bar  {\yng(2)}} \oplus {\bar  {\yng(1,1)}}  $     & $N^2  \cdot  {\underline {1}}  $     &   $ - (N+4) $    & $-1$ \\[1.8ex]
&  $ \eta_{a_2}$      &   $ 4  \cdot \Yvcentermath1  {\bar  {\yng(1)}}   $     & $N \cdot \Yvcentermath1 {\yng(1)}  $     &   $ - \frac{N+4}{2}  $   & $-1$  \\[1.8ex]
   \hline 
   $ \phantom{{\bar{ \bar  {\bar  {\yng(1,1)}}}}}\!  \! \!\! \! \!  \!\!\!$  {\rm IR}&      $ {\cal B}_{[a_1  b_1]}$      &  $\Yvcentermath1 {\bar  {\yng(1,1)}}   $         &  $  \frac{N(N-1)}{2} \cdot  {\underline {1}} $        &    $   -(N+4) $     & $-1$ \\[2ex]
       &   $ {\cal B}_{[a_1 b_2]}$      &  $   4 \cdot \Yvcentermath1 {\bar  {\yng(1)}}   $         &  $N \cdot \Yvcentermath1 {\yng(1)}  $        &    $ - \frac{N+4}{2}$   & $-1$   \\[1.8ex]
\hline
\end{tabular}  
  \caption{\footnotesize   Color-flavor locked phase in the $\psi\eta$ model.  
  $a_1$ or $b_1$  stand for  $1,2,\ldots, N$,   $a_2$ or $b_2$ the rest of the flavor indices, $N+1, \ldots, N+4$.   The fermion parity $({\mathbb{Z}}_{2})_F$  acts as  $\psi \to -\psi$, $\eta\to -\eta$.
   }\label{SimplestBis}
}
\end{table}
  The low--energy, massless degrees of freedom are  
    $\tfrac{N^2+7N}{2}$ massless baryons,
    \be       {\cal B}_{[a b]} =   \psi^{ij} \eta_{i a}  \eta_{j b}\;, \qquad   a,b = 1,2,\ldots N+4\;, 
    \ee
  together with  $8N+1$  Nambu-Goldstone (NG) bosons.

Let us recall briefly the role the discrete $({\mathbb{Z}}_{2})_F$  symmetry of  Table~\ref{SimplestBis}  played in 
 constraining the infrared dynamics of these models. 
As shown in \cite{BKL2,BKL4,BKLZ2, corfu},  the  fermion $({\mathbb{Z}}_{2})_F$ symmetry suffers from various mixed  anomalies such as $ [({\mathbb Z}_{2})_{F}]- [\mathbb{Z}_N]^2 $  and  $ [({\mathbb Z}_{2})_{F}]- [\mathbb{Z}_{N+4}]^2$,   in $N={\rm even}$  models, where  $\mathbb{Z}_N$ and  $\mathbb{Z}_{N+4}$ are 1--form 
(``generalized'') symmetries \cite{Seiberg, KapSei,AhaSeiTac,GKSW,GKKS,ShiYon,TanKikMisSak,Komargodski:2017smk,AnbPop1,AnbPop2,Tanizaki}. 
These anomalies fail to be matched by massless ``baryons''  in a hypothetical  confining phase with no global symmetry breaking,  sometimes discussed  in the literature.  Such a phase  therefore cannot be dynamically realized in the IR. No problem arises in the dynamical Higgs phase assumed here:
 the condensates $\brc \psi \eta \ckt$   (\ref{VEV1}) indeed breaks spontaneously both the global 0--form $U(1)$ and  the global 1--form    $\mathbb{Z}_N$   color center   (or  the flavor $\mathbb{Z}_{N+4}$  center) symmetries, thus ``matching'' the impossibility of gauging these center symmetries together with  $({\mathbb{Z}}_{2})_F$.

      By taking into account the $\chi$ fields, decomposed as a direct sum over irreducible representations (irreps) of $SU(N)^{\prime} \times SU(4)$,      and renaming the baryons as
        \be    {\cal B}_{[a_1  b_1]} \to \lambda_1\,, \qquad    {\cal B}_{[a_1 b_2]} \to \lambda_2\,, \ee   
        one has the  low--energy  gauge groups and  the  fermions  of this  model,   in Table~\ref{LE1}.      Note that both   $SU(N)^{\prime} $ and  $ SU(4) $ are anomaly--free and $U(1)^{\prime}$ is free of the ABJ  and   $[U(1)^{\prime}]^3,  [U(1)^{\prime}]-[{\rm gravity}]^2$  anomalies. 
         \begin{table}[h!t]
{
  \centering 
  \begin{tabular}{|c|c |c|c|  }
\hline    
${}_{\rm fields} \backslash {}^{\rm groups}$    &  $SU(N)^{\prime} $    &  $ SU(4) $     &   $  U(1)^{\prime}   $   \\
    \hline 
   &&&     \\[-2ex]
        $\lambda_1$      &  $\Yvcentermath1 {\bar  {\yng(1,1)}}   $         &  $ {\underline {1}} $        &    $   -(N+4) $      \\[1.6ex]
          $\lambda_2$      &  $\Yvcentermath1   {\bar  {\yng(1)}}   $         &  $\Yvcentermath1 {\yng(1)}  $        &    $ - \frac{N+4}{2}$      \\[1.6ex]
          $\chi_1$      &  $\Yvcentermath1    {{\yng(1,1)}}     $         &  $ {\underline {1}}  $        &    $N+4$     \\[1.6ex]
         $\chi_2$      &  $\Yvcentermath1    {\yng(1)}  $         &  $ \Yvcentermath1 \bar{\yng(1)}  $        &    $\frac{N+4}{2}$      \\[1.6ex]
         $\chi_3$      &  $  {\underline {1}} $         &  $\Yvcentermath1 {\yng(1,1)}  $        &    $0$     \\[1.6ex]
      \hline
\end{tabular}  
  \caption{\footnotesize   The  spectrum of the massless fermions in the model of Table~\ref{Simplest} at mass scale below  $\Lambda_1$, where
  the $SU(N)$ gauge interactions are assumed to have become strongly--coupled.    
   }\label{LE1}
}
\end{table}

  $  U(1)^{\prime}   $ is an anomaly-free, unbroken   combination of 
     a  broken subgroup of  $SU(N+4)$, generated by
      \be    \eta \to    \rme^{  \im Q  \beta }   \eta\;,      \qquad     Q=  \left(\begin{array}{cc}  4 \, \mathbbm{1}_N & 0 \\0 & - N \, {\mathbbm 1}_4\end{array}\right)
    \label{broken}    \ee
      and $U(1)$ of Table \ref{Simplest}, 
    \be
U(1): \ \psi \to \rme^{\im (N+4)\alpha}\psi\;, \quad  \eta \to \rme^{-\im (N+2)\alpha}\eta\;,  \quad   \chi \to \rme^{\im N \alpha}\chi\;. \label{upe0} 
\ee  
Requiring that the vacuum expectation value (VEV) (\ref{VEV1}) leaves it unbroken gives the condition 
\be  U(1)^{\prime}  \ :  \qquad       2 \alpha +  4 \beta=0\;,  \qquad  \beta=  - \frac{\alpha}{2} \;. 
\ee

The color-flavor diagonal  $SU(N)^{\prime}$, which is a global symmetry in the limit    $g_2 \equiv  g_{SU(N+4)} \to 0$,  is actually a local, gauge symmetry 
for   $g_2 \ne 0$.   Its coupling constant $g_{SU(N)}^{\prime}$  is a combination,  
\be  
g^{\prime} \equiv  \frac{g_1 g_2}{ \sqrt{g_1^2+ g_2^2}}\;, \qquad  g_1 \equiv  g_{SU(N)}\;, \quad    g_2 \equiv  g_{SU(N+4)}  \;,  
\label{mixing00}  
\ee
(see Appendix~\ref{mixing}).    
At that mass scale $\Lambda_1$, 
$g_2 \equiv g_{SU(N+4)}$ is assumed to be  weakly-coupled,  
\be   g_2(\Lambda_1)   \ll   g_{1}(\Lambda_1)     \;.     \label{hierarchy}    \ee
  We see that just   below the scale $\Lambda_1$   the surviving $SU(N)^{\prime}$  gauge interactions 
are  actually weakly-coupled,  with   
\be   g^{\prime} \equiv   g_{SU(N)}^{\prime}(\Lambda_1)    \simeq  g_2  \;, \label{similar} 
\ee
 due to the hierarchy (\ref{mixing00}), (\ref{hierarchy}). {Thus, the part of $SU(N+4)$ that becomes strong and is locked to $SU(N)$ survives, but in a weak form.}

The (first term of the)  beta functions of  $SU(N)^{\prime}$  and  $SU(4)$   below the mass scale $\Lambda_1$ are:
\begin{align}     \frac{d g^{\prime} }{d \log \mu  }   & \equiv     \beta(g^{\prime})  = 
   -  \frac{9 N -4}{3} \,   \frac{(g^{\prime})^3}{16\pi^2} + o((g^{\prime})^5) \equiv    b_0^{\prime}  \, \frac{(g^{\prime})^3}{16\pi^2}  + o((g^{\prime})^5) \;,  \nonumber \\
      \frac{d g_2 }{d \log \mu }     &\equiv      \beta(g_2)  = 
      - \frac{42 - 2 N}{3}  \,  \frac{g_2^3}{16\pi^2}  + o(g_2^5) \equiv    b_0^{(2)}    \,  \frac{g_2^3}{16\pi^2}    + o(g_2^5)  \;. \label{beta0}  
\end{align}
Thus  $SU(N)^{\prime}$ and $SU(4)$ are both asymptotically-free,  as long as $N <  21  $, which we assume to be the case below.  

   The system at the scale below $\Lambda_1$ also contains  $8N+1$ pseudo NG bosons  in the case $SU(N+4)$ is a global symmetry. 
The  $8N$    (would-be) NG bosons, which are not absorbed by the $SU(N)$ gauge bosons,  correspond to the nondiagonal generators of $SU(N+4)$ 
\be      T^{i k}\;, \quad T^{k i}\;,  \qquad  i=1,2,\ldots, N\;, \quad  k= N+1, \ldots N+4\;, 
\ee    
i.e., in $({\underline N}, {\underline 4})$ of   $SU(N)^{\prime}\times SU(4)$,  and absorbed by the associated $SU(N+4)$ gauge bosons. 
The last NG boson corresponds to the $SU(N+4)$ generator  (\ref{broken}) and is a singlet      $({\underline 1}, {\underline 1})$ of   $SU(N)^{\prime}\times SU(4)$.  This NG boson 
gets absorbed  by the corresponding  broken  $SU(N+4)$ gauge boson.

\begin{figure}
\begin{center}
\includegraphics[width=5.5 in]{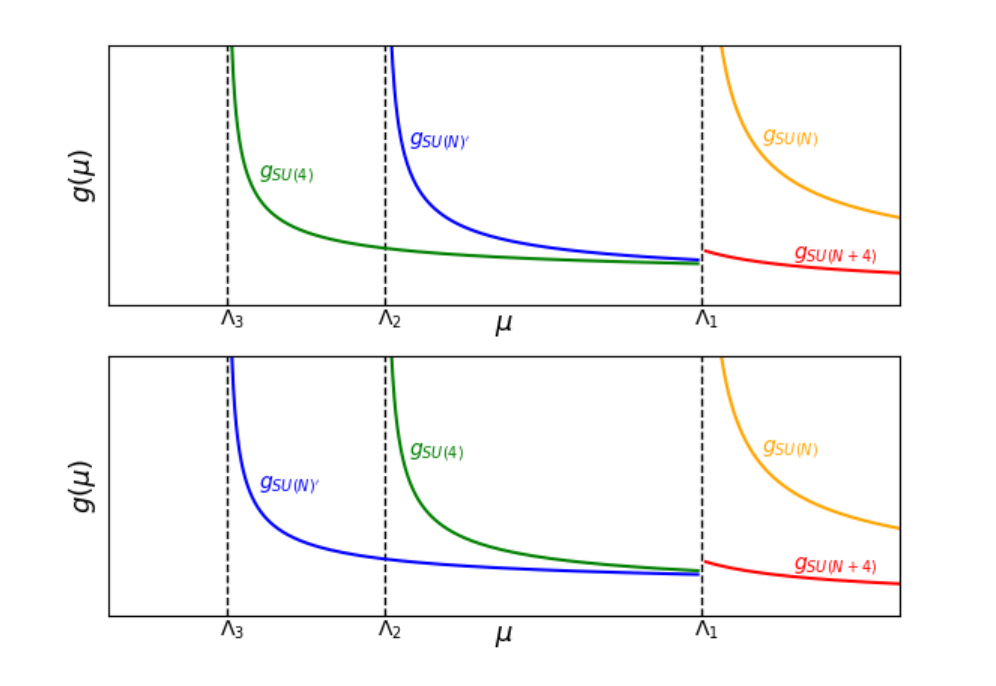}
\caption{\footnotesize   The RG flow of the coupling constants in the  $SU(N)-SU(N+4)$  model of Table.~\ref{Simplest}.  The top  picture corresponds to the  theory with $N \ge   5 $, the lower corresponds to $N  \le    4$.  }
\label{modello1}
\end{center}
\end{figure}

Below $\Lambda_1$  both $SU(N)^{\prime}$ and $SU(4)$  gauge interactions grow stronger towards the IR. Which of them becomes strongly coupled first   depends on the value of $N$, as their values just below $\Lambda_1$  are  close to each other (see (\ref{similar})).   The RG flow 
of the coupling constants in the  $SU(N)-SU(N+4)$  model  of this section   is illustrated  in Fig.~\ref{modello1} for  two possible cases, discussed in  Sec.~\ref{scenario1} and Sec.~\ref{scenario2}. 

\subsubsection{Large $N$  \label{scenario1}}
 Taking the value of $b_0$  (\ref{beta0})  as a rough indication,  
one may conclude that for a relatively large $N$  ($N \ge 5$)   $SU(N)^{\prime}$  couplings become stronger first, flowing towards the IR,  at some mass scale $\Lambda_2$,
\be (\Lambda_2)^2  =   (\Lambda_1)^2   \,   e^{16\pi^2 / b^{\prime}_0  \, g^{\prime}(\Lambda_1)^2  }    \ll    (\Lambda_1)^2\;.
\ee
The $SU(N)^{\prime}$ theory is a QCD,  with a Dirac pair ($\lambda_1, \chi_1)$ in the anti anti-symmetric tensor representation and $N_f=4$ flavors of the fundamental ``quarks'',  ($\lambda_2, \chi_2)$,  which are all expected to condense in pairs and get mass.  
The global  $U(1)^{\prime}$ symmetry is unbroken, thus no NG bosons appear, and as the    $\chi_3$ has no charge with respect to it.

At mass scales below  $\Lambda_2$,   there remains unbroken $SU(4)$ gauge system with  a self-adjoint  fermion  $\chi_3$:  it will evolve further towards the IR and it is expected to become strongly coupled at mass scale $\Lambda_3$,  
$\Lambda_3 \ll \Lambda_2 \ll \Lambda_1$, goes into confinement phase,   forming  the condensate
\be     \brc \chi_3  \chi_3 \ckt\;,\ee 
and produces mesons  $ \chi_3  \chi_3$  of mass of the order of    $ \Lambda_3$, 
\be   \Lambda_3 \ll \Lambda_2 \ll \Lambda_1\;.
\ee

\subsubsection{Small $N$  \label{scenario2}}

If $SU(4)$ gets stronger first towards the IR  ($N \le 4$),   the system will go into the $SU(4)$  confinement phase at some scale  $\Lambda_2
\ll   \Lambda_1$, 
with condensates
\be       \brc \lambda_2  \chi_2 \ckt\;, \qquad   \brc \chi_3  \chi_3 \ckt\;
\ee
of the order of $(\Lambda_2)^3$,   forming.  
These fermions get dynamical masses of the order of   $\Lambda_2$ and decouple from the system at lower energies.  
At mass scales below  $\Lambda_2$  the system is an  $SU(N)^{\prime}$   gauge theory which looks like a one-family (antisymmetric) tensor QCD
with fermions $\lambda_1, \chi_1$  (see Table~\ref{LE1}).  It will get strongly coupled at $\Lambda_3$ ($\Lambda_3 \ll \Lambda_2  \ll \Lambda_1$),  confines, and  produces  mesons and baryons.

\subsection{$SU(N+4)$ getting strong first}

Let us consider now  the case $SU(N+4)$ coupling gets strong first, towards the IR, at some scale $\Lambda$,  in our $SU(N)-SU(N+4)$  model
of Table~\ref{Simplest}.  
The $SU(N)$ coupling constant  $g_1 \equiv g_{SU(N) }$ is now  assumed to be small at $\Lambda$. 
The dynamics of the system will then basically be the same as in what was known as the Georgi-Glashow model  (the ``$\chi\eta$ model''), studied in \cite{ADS,BKS,BKL4,BKL5,BKLReview,BKLproc,BKLZ2,corfu, BKLO1}.
A color-flavor locked bifermion condensate 
\be   \brc   \chi^{[ab]} {\eta}_{i\, b}   \ckt  = \const   \, (\Lambda_2)^3 \delta^a_{i}   \;, \qquad i, a,b=1,2,\ldots, N\; \label{cfl01}
\ee
is expected to form,  based on considerations of mixed anomalies, as in the case of the  Bars-Yankielowicz models.  
Such a VEV leaves  a diagonal   $SU(N)^{\prime} \subset   SU(N) \times SU(N+4)  $ as well as  $SU(4)\subset SU(N+4)$ 
as the unbroken gauge  groups.  The massless fermions surviving the condensate formation (\ref{cfl01})  have been enlisted in Appendix D of 
\cite{BKLReview},  see Table~\ref{SimplestAgain2} below\footnote{Below,  in  adapting the results  found in  \cite{BKLReview} to   present model of Table~\ref{Simplest},  we make a shift  $N \to N+4$, and take the complex conjugation of all the representations.  Lastly, the indices 
are appropriately renamed so as to agree with those used in previous sections in the present paper.}.

The massless baryons 
\be      B_{\{ij\}} = \chi^{[ab]} \,  {\eta}_{a\, i}   {\eta}_{b\, j}   \;, \qquad   i,j=1,2,\ldots N\;, 
\ee
are believed to form, and 
 saturate all the anomalies associated with $SU(N)^{\prime}   \times  U(1)^{\prime}$.  This is manifestly seen in Table~\ref{SimplestAgain2}: it is an example of the ``natural anomaly matching''  \cite{BKLO1}\footnote{
 After the fermions participating in the condensate pairwise get Dirac masses and decouple,  
the remaining massless fermions in the UV and in the IR are identical, with respect to the unbroken groups.}.

 \begin{table}[h!t]
  \centering 
  \begin{tabular}{|c|c|c|c|c|  }
\hline
   & ${}_{\rm fields} \backslash {}^{\rm groups} $    &   $SU(N)^{\prime}$      &   $ U(1)^{\prime} $     &  $SU(4) $     \\
 \hline
 &&&& \\ [-2ex]
   {\rm UV}&  $\chi^{a_1 b_1}$     &    $\Yvcentermath1  {{ \yng(1,1)}}   $    & $N$   &   ${\underline {1}}$ \\[1.6ex]
 &  $\chi^{a_1 a_2}$   &    $\Yvcentermath1   { { \yng(1)}} $    & $\frac{N}{2}$   &   $\Yvcentermath1 { { \yng(1)}}   $     \\[1.6ex]
 &$\chi^{a_2 b_2}$   &    $  {\underline {1}} $    & $0$    &   $\Yvcentermath1 { { \yng(1,1)}}   $     \\[1.6ex]
& $ {\eta}_{a_1,i}$          & $\Yvcentermath1 {\bar {\yng(2)}} \oplus {\bar {\yng(1,1)}}$     &   $ - N $    &   $ {\underline {1}}  $  \\[1.6ex]
 &   $ {\eta}_{a_2,i}$         &   $\Yvcentermath1 {\bar {\yng(1)}}  $     &   $ - \frac{N}{2} $     &   $\Yvcentermath1   {\bar {\yng(1)}} $     \\[1.6ex]
   \hline 
     \phantom{\huge i}$ \! \!\!\!\!$  {\rm IR}&     $ B_{\{ij\}}$        &  $\Yvcentermath1 {\bar  {\yng(2)}}$        &    $ - N $     &  $ {\underline {1}}  $      \\[1.6ex]
          \phantom{\huge i}$ \! \!\!\!\!$  &     $ \chi  $        &  ${\underline {1}}$        &    $ 0  $      &      $\Yvcentermath1 { { \yng(1,1)}}   $    \\[1.6ex]
\hline
\end{tabular}
  \caption{\footnotesize  Color-flavor locking  in the $``\chi\eta$'' (generalized Georgi-Glashow) model.    The color index $a_1$, $b_1$ and $i$    run up to $N$ and the rest,  $a_2$ or $b_2$, covers $N+1,\ldots, N+4$.}\label{SimplestAgain2}
\end{table}

\begin{table}[h!t]
  \centering 
  \begin{tabular}{|c|c |c|c|  }
\hline
${}_{\rm fields} \backslash {}^{\rm groups} $     &   $SU(N)^{\prime}$      &   $ U(1)^{\prime} $     &  $SU(4) $     \\
      \hline 
      &&& \\ [-2ex]
              $ \kappa  $        &  $\Yvcentermath1 {\bar  {\yng(2)}}$        &    $ - N $     &  $ {\underline {1}} $      \\[1.6ex]
           \phantom{\huge i}$ \! \!\!\!\!$           $\psi$ 
   &      $ \Yvcentermath1 { \yng(2)} $  &     $ N $    &    $ {\underline {1}} $   \\[1.6ex]
      \phantom{\huge i}$ \! \!\!\!\!$         $\chi$   &   ${\underline {1}}$       &   $0$      & $\Yvcentermath1 {\yng(1,1)}  $    \\[1.6ex]
\hline
\end{tabular}
  \caption{\footnotesize   The low--energy spectrum of the  $SU(N)-SU(N+4)$  model
of Table~\ref{Simplest},  below the scale $\Lambda_2$  at which the $SU(N+4)$ gauge interactions get  strong.  
   }\label{Simplest33}
\end{table}

\begin{figure}[H]
\begin{center}
\includegraphics[width=4in]{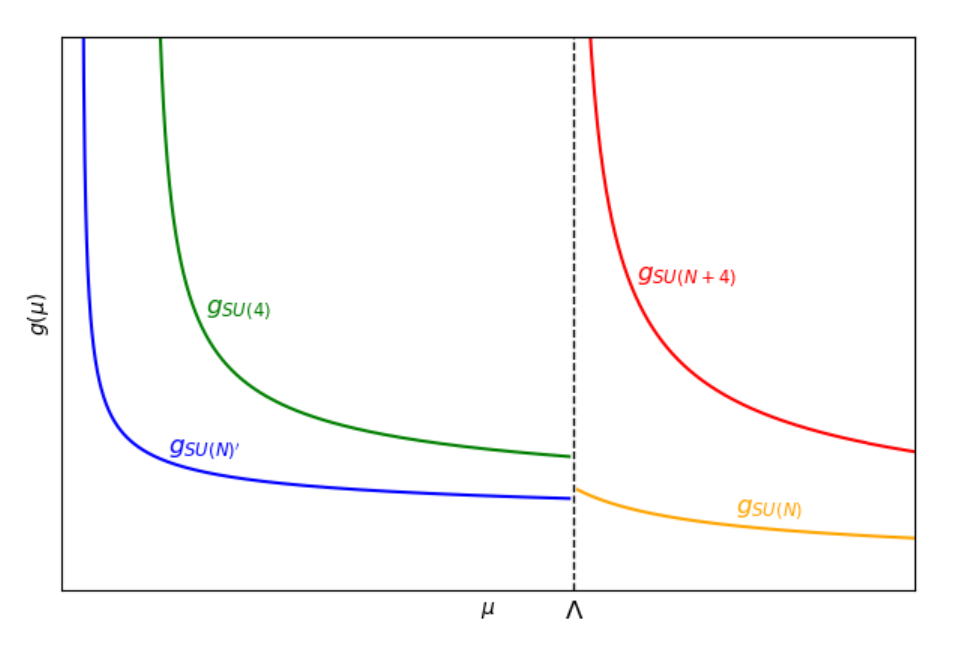}
\caption{ \footnotesize The RG flow of the system Table~\ref{Simplest}, where  the  $SU(N+4)$ interactions become strong first towards the IR at the mass scale $\Lambda$, at which the other, $SU(N)$  gauge interactions are assumed to be  weak, $ g_{SU(N)} \ll 1$.  Just below  $\Lambda$, both the diagonal   $SU(N)^{\prime} \subset   SU(N) \times SU(N+4)$ and   $SU(4)\subset SU(N+4)$  are weakly-coupled.   They  are both asymptotically-free and evolve independently.  
  }
\label{SUN+4}
\end{center}
\end{figure}

In the present model,  the ``color-flavor locked''  $SU(N)^{\prime}$  gauge group is actually a local gauge group, as the original $SU(N)$ group 
is local, though weakly-coupled.   By taking into account the $\psi$ field, and renaming  
\be   B_{\{ij\}}  \to    \kappa\,, \ee   the infrared spectrum of the system below    the mass scale $\Lambda$  is summarized in Table~\ref{Simplest33}. It is just a one-flavor (symmetric) tensorial QCD, together with a
decoupled $SU(4)$ sector with a single fermion  $\chi$   in the self-adjoint antisymmetric tensor representation  ($\chi^{a_2 b_2}$  of Table~\ref{SimplestAgain2}).  We assume that 
$SU(4)$ confines and the condensate
\be  \brc  \chi \chi \ckt \ne 0 \;, 
\ee
forms and $\chi$ acquires mass dynamically.  So do the tensor ``quarks''  $\psi$, $\kappa$, confined by the  $SU(N)^{\prime}$  forces. 
The RG flow of the coupling constants  in this case is illustrated in Fig.~\ref{SUN+4}.  

\section{The quiver     model with multiple $SU(N)$'s \label{quiver}}

\subsection{$SU(N)^3$ quiver model}

The second model we consider is a quiver   $SU(N)^3$    model,  an $SU(N)_1 \times SU(N)_2 \times SU(N)_3$ gauge theory with fermions 
in the following  quiver--type bi--fundamental representations (Table~\ref{Simplest4}).

Let us assume that one of the $SU(N)$ factors gets strongly   coupled first.   As the three  $SU(N)$'s appear symmetrically, we may assume without loss of generality that 
\be    \Lambda_{1}  \gg     \Lambda_2,  \Lambda_3\;.   
\ee
That is, we assume that   $SU(N)_2$   and  $SU(N)_3$   are weakly-coupled on the scale $ \Lambda_{1} $ where $SU(N)_1$ is strongly   coupled.
That is   
\be    g_1(\mu)  \gg    g_2(\mu) , \, g_3(\mu)      
\ee
near    $\mu \simeq  \Lambda_1$, where $ g_1(\mu) \ge 1$. 

\begin{table}[h!t]
  \centering 
  \begin{tabular}{|c|c |c|c|  }
\hline
${}_{\rm fields} \backslash {}^{\rm gauge \, group}   $     &   $SU(N)_1$      &   $SU(N)_2$      &  $SU(N)_3$     \\
      \hline   
      &&& \\ [-2ex]
            $ \psi  $        &  $\Yvcentermath1 {{\yng(1)}}$        &   $\Yvcentermath1 {\bar {\yng(1)}}$     &  $ {\underline {1}}    $      \\[1.6ex]
          
           $\kappa$ 
   &      $\Yvcentermath1 {\bar {\yng(1)}}$     &      $ {\underline {1}}    $     &    $\Yvcentermath1 {{\yng(1)}}$     \\[1.6ex]
      $\lambda$   &  $ {\underline {1}}    $        &  $\Yvcentermath1 {{\yng(1)}}$        & $\Yvcentermath1 {\bar {\yng(1)}}$    \\[1ex]
\hline
\end{tabular}
  \caption{\footnotesize    The  quiver  $SU(N)^3$    model with fermions $\psi, \kappa, \lambda$.  There are no anomaly-free combinations of classical $U(1)$ symmetries.    }\label{Simplest4}
\end{table}

Neglecting first the weak   $SU(N)_2$   and  $SU(N)_3$ interactions, 
i.e., by setting  $g_2=g_3=0$,   
the system is just a $N_f=N$ flavored   $SU(N)$ QCD, together with a free
(spectator) fermions
$\lambda$ that takes care of the $SU(N)_2$   and  $SU(N)_3$  anomalies.   We know that a ``quark condensate''
\be     \brc  \psi^i_a  \kappa^k_i \ckt  =  \const \,   \delta_a^k \, (\Lambda_1)^3   \;,  \qquad   i, a, k =1,2, \ldots, N\;         \label{QC}
\ee
 forms,   and the $SU(N)_2  \times SU(N)_3$  flavor symmetry is broken to  the diagonal  
\be       SU(N)_0 \subset     SU(N)_2  \times SU(N)_3\;. 
\ee
The fermions $\psi$ and $\kappa$ become massive, acquiring dynamical masses of the order of  $\Lambda_1$.  They form  $SU(N)_1$-singlet mesons and baryons of masses of the order of  $\Lambda_1$.  

The  orthogonal,  axial $SU(N)_A \subset  SU(N)_2  \times SU(N)_3$  is broken spontaneously and produces  massless  $N^2-1$ NG bosons. 
In the presence of the weak $SU(N)_2  \times SU(N)_3$  gauge interactions,   
 these  $N^2-1$ massless  ``pions''  get absorbed by the axial $SU(N)_A $ gauge bosons, giving them masses of the order of $ g^{\prime}  \Lambda_1$,   $ g^{\prime}  \equiv \sqrt{g_2^2+ g_3^2}$. 
The decomposition and gauge-field mixing is discussed in Appendix \ref{mixing2}.

At mass scales much lower than  $ g^{\prime} \Lambda_1 ( \ll \Lambda_1)$,   the massless degrees of freedom are 
 the  $SU(N)_0$  gauge bosons and the fermions $\lambda$ in the adjoint representation of $SU(N)_0$.   
 Note that in the decomposition of  $\lambda$   in  $  SU(N)_0  \subset  SU(N)_2  \times SU(N)_3$,   
\be      {\underline N} \otimes   {\underline N}^* =       {\underline {N^2-1} } \oplus    {\underline 1} \;,  
\ee
the singlet part decouples from the $SU(N)_0$ gauge bosons, i.e. it is a free, massless fermion, $\lambda_0$.

At mass scales  below   $ g^{\prime} \Lambda_1 $
the massless fermions $\lambda$  are seen to be coupled   as    
 \be        {\cal D}_{\mu}  \lambda =    \de_{\mu}  \lambda  -   i g_0   \, [ A_{\mu}^{(0)},  \lambda]   \;,      
 \ee
 to the gauge bosons  $A_{\mu}^{(0)}$ of the unbroken gauge group    $SU(N)_0$,     
 where
 \be   g_0 \equiv    \frac{g_2  g_3}{g^{\prime} }=   \frac{g_2  g_3}{\sqrt{g_2^2+ g_3^2}}   \;.  \label{SYMcoupling}
 \ee
 Note that it is smaller than both  $g_2$ and $g_3$.    This formula also shows that  the evolution towards the IR of the system   below the mass scale 
 $ g_{2,3}\Lambda_1 $  is essentially independent of the detailed relation between   $\Lambda_2$ and $ \Lambda_3$, or  the relative strengths  between the small  coupling constants   $g_2(\mu)$ and   $g_3(\mu)$  near    $\mu \sim   \Lambda_1$.

\begin{figure}[H]
\begin{center}
\includegraphics[width=5in]{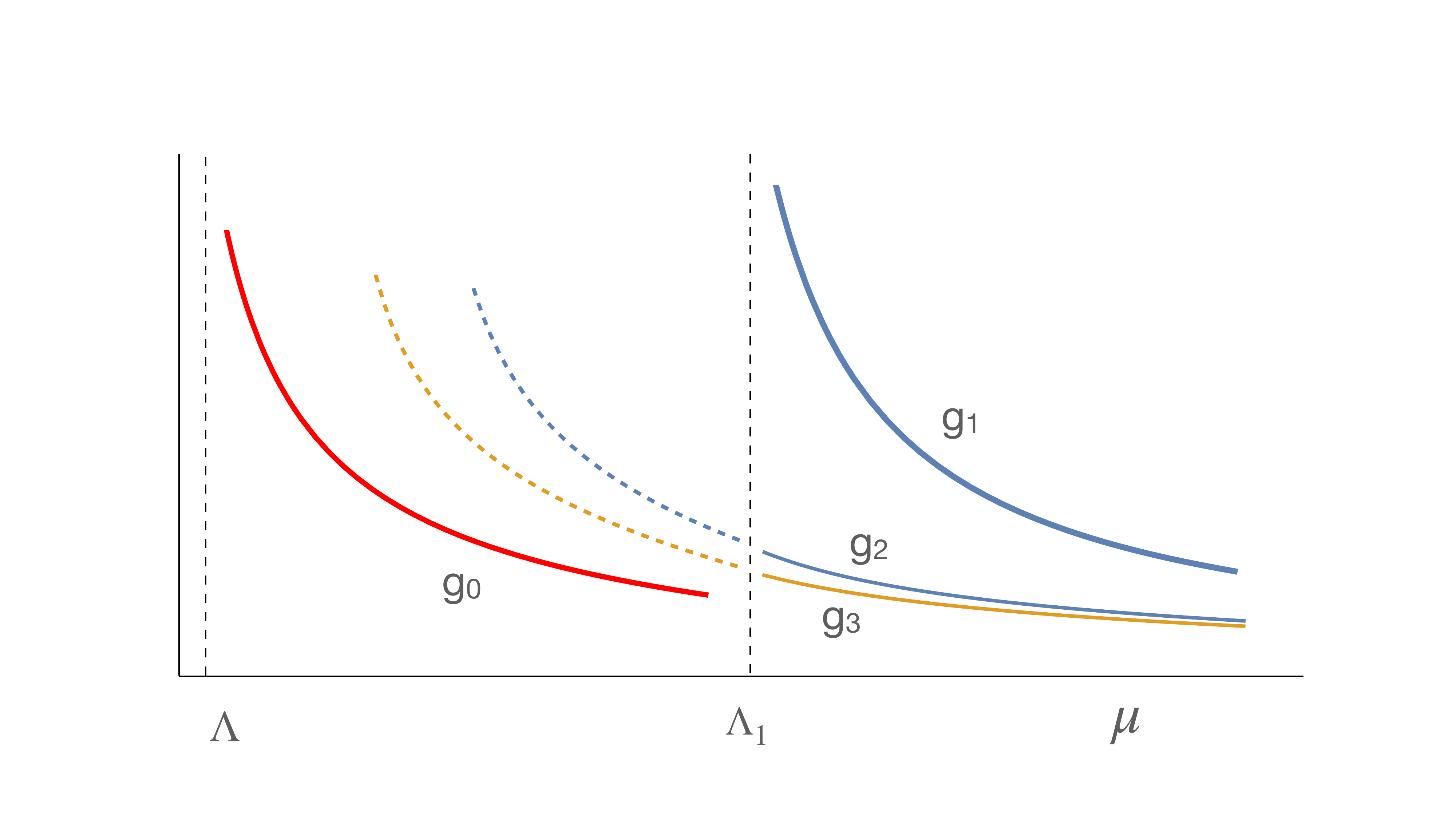}
\caption{ \footnotesize  A schematic view of the RG flow of the coupling constants in the quiver $SU(N)^3$ model.  $g_{1,2,3}(\mu)$ represent the coupling constants of $SU(N)_1$, $SU(N)_2$ and $SU(N)_3$, respectively, as functions of the mass scale $\mu$.  At $\mu\simeq \Lambda_1$  the $SU(N)_1$ interactions become strong, and 
assumed to confine and form the $\psi\kappa$ condensate, (\ref{QC}).  The axial combination of $SU(N)_2$ and $SU(N)_3$ gets spontaneously broken and the corresponding gauge bosons acquire masses $\sim g_{2,3} \, \Lambda_1$.  Below  $g^{\prime}  \Lambda_1$  the system becomes effectively an ${\cal N}=1$
supersymmetric  $SU(N)$  Yang-Mills theory.   }
\label{RGquiver}
\end{center}
\end{figure}

The system is now  ${\cal N}=1$ supersymmetric pure Yang-Mills  (SYM) theory, thoroughly  studied earlier \cite{Amati1984,Novikov:1982px,Affleck:1983mk,Novikov:1985ic,Amati:1988ft,Davies:1999uw,Finnell:1995dr,Konishi2003}. The fermion $\lambda$ plays the role of the gauge fermion (the ``gluino'').    What is the RG-invariant mass scale $\Lambda$ of this emergent  $SU(N)$ SYM?  In terms of the  $SU(N)_0$  coupling constant   (\ref{SYMcoupling})   below the scale $\Lambda_1$,  it is given by  
 \be \Lambda^2  =   (\Lambda_1)^2    \,   e^{16\pi^2 /  b_0  \, g_0(\Lambda_1)^2  }    \ll    (\Lambda_1)^2\;,
\ee
where 
\be    b_0 =  -  3N 
\ee
is the well-known (first coefficient of the)  beta function of the $SU(N)$ SYM theory.   The RG flow of this model  is  skeched in Fig.~\ref{RGquiver}.

There are $N$ vacua,  in which  the gluino condensate takes the known exact values \cite{Amati1984,Novikov:1982px,Affleck:1983mk,Novikov:1985ic,Amati:1988ft,Davies:1999uw,Finnell:1995dr,Konishi2003}
\be       \left\brc   \frac{\Tr \lambda \lambda}{16\pi^2} \right\ckt=      \Lambda^3  \, e^{2\pi i k/N}\;, \qquad k=1,2,\ldots, N\;. 
\ee
In each vacuum, at  mass scales much  lower than $\sim \Lambda$, the world consists of just the free fermion(s), $\lambda_0$. 

\subsection{$SU(N)^n$ quiver model}

\begin{figure}[H]
\begin{center}
\includegraphics[width=5.5in]{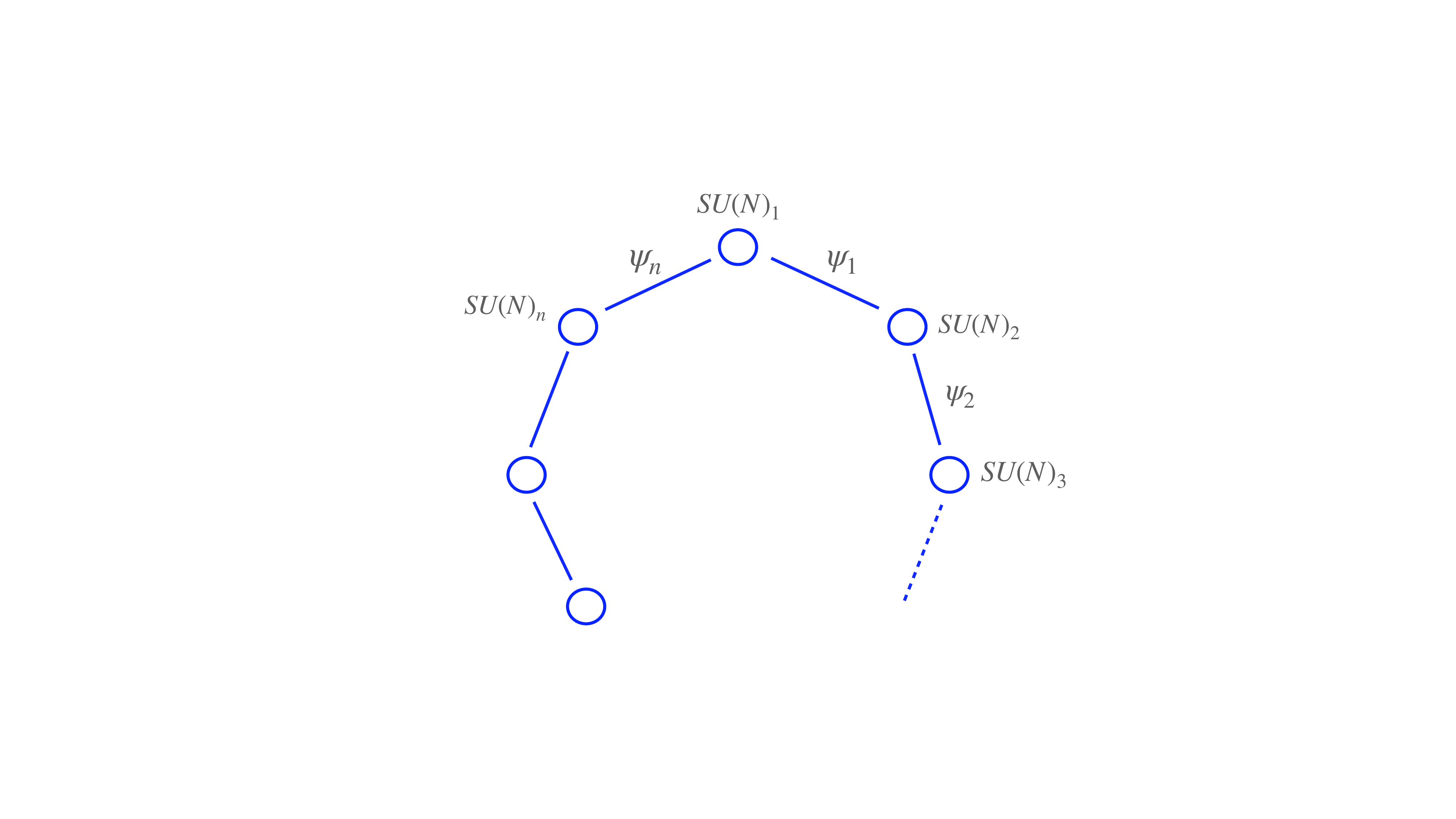}  
\end{center}
\caption{ \footnotesize   a $SU(N)^n$ quiver model. }
\label{quivers}
\end{figure}

A question comes into one's mind naturally.   What happens in the   $SU(N)^n \equiv SU(N)_1 \times SU(N)_2 \times \ldots   \times  SU(N)_n$
model with fermions in a bi-fundamental representation of the two adjacent $SU(N)$ gauge group factors  (see Fig.~\ref{quivers})?  
By repeating the argument above for $SU(N)^3$ quiver model,  it is easy to convince oneself that when one of the $SU(N)$ factors gets strongly coupled and goes into confinement phase,  the  two ``adjacent''  $SU(N)$ factors  combine into ``axial'', broken $SU(N)$, and unbroken diagonal 
$SU(N)$ gauge groups.  The net  effect is  the chain evolution towards the IR,  
\be 
SU(N)^n \to SU(N)^{n-2}  \to SU(N)^{n-4}  \to \ldots\;.  
\ee
where at each step the model is a quiver gauge theory with each fermion in the bifundamental representation with respect to the two adjacent 
gauge groups.  When $n$ is odd,  the system eventually  ends up with  the $SU(N)^3$ quiver model discussed above, and  in the extreme infrared it evolves as  a pure ${\cal N}=1$ supersymmetric Yang-Mills system, plus a decoupled free massless fermion. 

When $n$ is even, the system ends up with an   $SU(N)_1 - SU(N)_2$ quiver model in the infrared limit.  When one of the $SU(N)$ factors, say,  
 $SU(N)_1$,  becomes strongly  coupled at the mass scale $\Lambda$, it confines and all fermions get a dynamical mass of the order of $\Lambda$, producing massive mesons and baryons of mass $\sim \Lambda$. Depending on the coupling  $g_{SU(N)_2}$  near $\Lambda$,  there may 
 appear somewhat lighter broken gauge bosons, of masses of the order of    
 \be     g_{SU(N)_2} \, \Lambda\;.  
 \ee

\section{The $SU(N)- Sp(6)-Sp(6)$  model} 
\label{sec:4}

The third model we consider is an   $SU(N) \times  Sp(6)_L \times  Sp(6)_R$ 
gauge theory\footnote{We use the terminology  $Sp(6)_L$  and $Sp(6)_R$,   having in mind an analogy of this system with an $SU(N)$   $N_f=6$  QCD,  with  $\psi=  \psi_L$ and  ${\tilde \psi} \equiv \psi_R^c$.   The choice of this model was inspired by attempts to construct a model of color-electroweak-family  unification  \cite{KuoNaka,Davighi}, based on the  $SU(4) \times  Sp(6)_L \times  Sp(6)_R$  group,  even though in the present work we make no attempt to construct a realistic model of the fundamental interactions. 
 }   with fermions shown in Table~\ref{Simplest3}.    
$N=2n$, $n \in \mathbb{N}$ is taken to be even to ensure the absence of the Witten anomaly \cite{Witten:1982fp}.  Both $Sp(6)_{L,R}$ are asymptotically-free for $N <   44$ :
\be  b_0[Sp(6)_{L,R}]= -  \frac{11 \cdot 4 -       N }{3}   <    0\;, \qquad   N <   44  \;, 
\ee 
whereas  $SU(N)$ is AF for  any $N \ge 2$: 
\be  b_0[SU(N)]=-    \frac{11 \cdot N -     2 \cdot 6 }{3}   <    0\;.
\ee 
  Again, the infrared spectra and the structure of the low-energy effective theories are considerably different,   depending on which gauge interactions get stronger first, flowing towards the IR. 
  
  \begin{table}[h!t]
  \centering
  \begin{tabular}{|c|c |c|c| }
\hline
$   {}_{\rm fields} \backslash {}^{\rm gauge \, group}$&  $SU(N) $   &   $ Sp(6)_L $ &  $ Sp(6)_R$     \\
 \hline     
     &     
       &       &          \\[-2ex]
       $\psi $     &     
     $ \Yvcentermath1 { \yng(1)} $   &   $  {\underline {6}} $    &  $  {\underline {1}} $       \\  [1.6ex]
  $ {\tilde \psi}  $    &   $ \Yvcentermath1  {\bar  {\yng(1)}}$        &  $  {\underline {1}} $   &  $   {\underline {6}}  $ 
    \\ [.5ex]
\hline
\end{tabular}  
  \caption{\footnotesize   The matter Weyl fermions in the $SU(N) \times  Sp(6)_L \times  Sp(6)_R$ model. }
   \label{Simplest3}
\end{table}

\subsection{$SU(N)$  more strongly coupled than $Sp(6)$ \label{SUvsSp}  }

Let us first consider the case  the  $SU(N)$ forces become strong at some scale $\Lambda$,
assuming that at that scale  $Sp(6)_{L,R}$ interactions are weakly-coupled,
\be      g_{Sp_L} \sim g_{Sp_R}  \ll   g_{SU(N)}\;.  
\ee
Below, we shall sometime denote   $g_{Sp_L},   g_{Sp_R}$  generically as   $g_{\rm weak}$.

In the limit
\be      g_{Sp_L} = g_{Sp_R} =0\;, \label{weak}
\ee
i.e.,   $g_{\rm weak}=0$,  
the system is just an $SU(N)$ QCD  with six flavors $N_F=6$ of quarks.   The global symmetry of the system would be  
\be     G_F=   SU(6)_L \times SU(6)_R \times U(1)_V \;,  \label{approximateS} 
\ee
 and   one expects that the quark condensate
\be      \brc \psi^i_a  {\tilde \psi}_i^b \ckt  =    \Lambda^3     \delta_a^b \;,\label{condens1}
\ee
forms and breaks the global symmetry as
\be     G_F \to     SU(6)_V \times   U(1)_V\;,  \label{pattern0}  
\ee
generating 
$   35   $
NG bosons of the broken  $SU(6)_A$ symmetry. 

In our system, the subgroup $Sp(6)_L \times  Sp(6)_R \subset  SU(6)_L \times SU(6)_R$ is weakly gauged.  This appears to present a typical but subtle vacuum alignment problem of strong interaction dynamics versus global symmetry and weak gauging. 
Actually, the situation looks more straightforward, once we assume that the effect of the strong $SU(N)$ dynamics,  (\ref{condens1}),  is not affected  substantially by the weak gauging of the  subgroup $ Sp(6)_L\times Sp(6)_R \subset SU(6)_L\times SU(6)_R$.  

The condensate  (\ref{condens1})  reads
 \be \brc \psi_1  {\tilde \psi}^1 \ckt  =\brc \psi_2  {\tilde \psi}^2 \ckt  =\brc \psi_3  {\tilde \psi}^3 \ckt  =\brc \psi_4  {\tilde \psi}^4 \ckt  =\brc \psi_5  {\tilde \psi}^5 \ckt  =\brc \psi_6  {\tilde \psi}^6 \ckt  = \Lambda^3\,. \label{condens2}
 \ee
 By appropriately identifying the six components of $\tilde \psi$ as 
 \be  \tilde \psi=   \Omega \,  \psi^{\prime}\;,  \qquad  \Omega =  \left(\begin{array}{cc}{\mathbf 0}_3 & {\mathbbm 1}_3 \\-{\mathbbm 1}_3 & {\mathbf 0}_3\end{array}\right)\;,  \qquad      \psi^{\prime} \sim   {\underline 6}   \label{condsp0}
 \ee
 the condensation  (\ref{condens1}), (\ref{condens2})   can be rewritten as 
 \be          \brc \psi  \Omega \psi^{\prime}  \ckt   \propto {\mathbbm 1}\;,    \label{condsp1}
 \ee
 i.e., in the singlet of the decomposition  in $Sp(6)$, 
 \be {\underline 6 }\otimes  {\underline 6 } =  {\underline {1}} \oplus  {\underline {14}} \oplus {\underline {21}}    \;.    \label{show}
\ee 
It means the weak gauge symmetry is broken as 
\be  Sp(6)_L \times Sp(6)_R  \to   Sp(6)_{\rm diag} \; .     \label{result} \ee

This implies that, among the $35$ would-be NG bosons of the $N_f=6$ QCD,   
$21$  corresponding to the broken axial $ Sp(6)_A$ symmetry  are absorbed by the corresponding gauge bosons giving them mass, $\sim g_{\rm weak} \Lambda$. This is  a dynamical Higgs phenomenon. 

The rest, $14$ of them, can naturally be assigned to the multiplet  ${\underline {14}}$ of the unbroken $Sp(6)_{\rm diag}$ symmetry.  
They will acquire a mass of the order of  $g_{\rm weak} \Lambda$.  Note that our system has no exact global symmetry,  but only an approximate global symmetry (\ref{approximateS}),  which is broken {\it  explicitly}  by the 
$Sp(6)_L \times Sp(6)_R$ gauging.  Hence exactly massless physical NG bosons cannot appear. 
The expected mass $g_{\rm weak} \Lambda$ of these scalar particles in ${\underline {14}}$  is due to the standard quadratic divergence diagram, but here the (physical) cutoff is provided by the fact that these are composite particles, pointlike only at the mass scale $\mu \ll \Lambda$\footnote{
This might be considered as a typical situation where the Coleman-Weinberg potential \cite{Coleman:1973jx} comes into play. The massless particles in ${\underline {14}}$, here, however, are would-be NG bosons. A vacuum instability (a nonvanishing VEV of these particles) would imply a modification of the  symmetry breaking pattern  (\ref{condens1}), (\ref{condsp1}) and   (\ref{result}), which cannot occur for arbitrarily weak  $Sp(6)_L \times Sp(6)_R$  couplings.  A study of the Coleman-Weinberg potential in a similar situation \cite{Piai}  appears to be consistent with this consideration.   }.

At the same time,  $SU(N)$  interactions confine all fermions participating in the bifermion condensate 
 (\ref{condens1}):   all $\psi$ and ${\tilde \psi}$ ($\psi^{\prime}$)  are confined and get Dirac (constituent) mass of the order of $\Lambda$.  
 Mesons and baryons,
 \be   \psi {\tilde {\psi}}\;, \qquad  \psi \psi \ldots \psi\;, \qquad   {\tilde {\psi}} {\tilde {\psi}} \ldots   {\tilde {\psi}}
 \ee
 of mass of the order of $\Lambda$ will appear,  but they are now in the  irreps of 
 $Sp(6)_{\rm diag}$,  rather than of $SU(6)_V$.  For weak $Sp(6)_L \times  Sp(6)_R$ gauging, however, they will remain in approximate multiplets
 of  $SU(6)_V$. 
 For instance, the  ``$\rho$'' mesons in   ${\underline {35}}$ 
 \be {\underline {6}} \otimes  {\underline {6}}^* =   {\underline {1}} \oplus  {\underline {35}}   \label{NGB}
 \ee
 will split as   ${\underline {14}}\oplus {\underline {21}}$ of   $Sp(6)_{\rm diag}$,  see (\ref{show}),
 by small mass differences of the order of    $g_{\rm weak} \Lambda$.    These massive mesons and baryons carry charges with respect to 
 $Sp(6)_{\rm diag}$ and coupled to the  $Sp(6)_{\rm diag}$ gauge bosons, accordingly.

Summarizing, the spectrum of the $SU(N)\times Sp(6)_L\times Sp(6)_R$ theory of Table \ref{Simplest3}
is as follows. 
\begin{description}
  \item[(i)]  Massive mesons and baryons of mass, $\sim \Lambda$,  in various irreps of    $Sp(6)_{\rm diag}   \subset  Sp(6)_L \times Sp(6)_R$.
  They are also in  approximate multiplets of $SU(6)_V$.   
  \item[(ii)]  The fermions $\psi, {\tilde \psi}$ are all confined and get dynamical, constituent masses of the order of $\Lambda$;
  \item[(iii)]    Massive gauge bosons  of  broken   $Sp(6)_{A}$  group, of masses of the order of  $g_{\rm weak} \Lambda$. 
      \item[(iv)]   Massless gauge bosons  of  unbroken   $Sp(6)_{\rm diag}$;
  \item[(v)]    Light  pseudo NG bosons of masses $\sim g_{\rm weak} \Lambda$, in  $\underline{14}$  of $Sp(6)_{\rm diag}$.
\end{description}

\begin{figure}
\begin{center}
\includegraphics[width=6in]{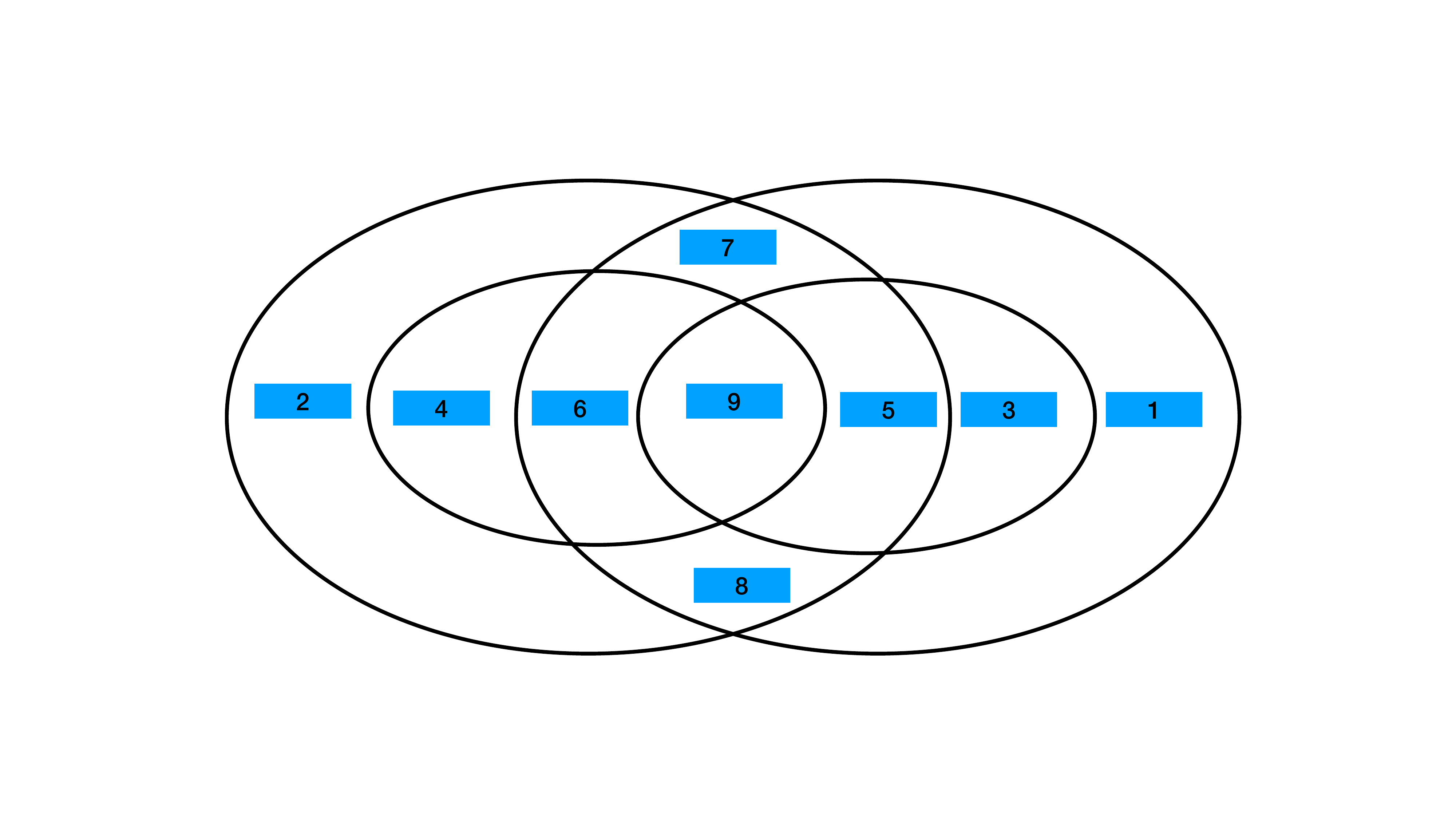}
\caption{\footnotesize   Symmetries acting on the fermions. The two large ellipses  including the sections (2467895) and  (1367895) represent
$SU(6)_L$ and $SU(6)_R$ transformations, respectively. The two smaller ellipses containing the sections  (469) and (359)  represent the subgroups 
$Sp(6)_L\subset SU(6)_L$ and $Sp(6)_R\subset SU(6)_R$, respectively.   The combined section (56789)  represents the intersection  $SU(6)_V=   SU(6)_L  \cap   SU(6)_R$, while   the section (9)  gives   $Sp(6)_{\rm diag} =   Sp(6)_L  \cap   Sp(6)_R$.   The section (5,6) represent some generators of $SU(6)_V$  which is weakly gauged but not part of  $Sp(6)_{\rm diag}$.  The section (7,8)  correspond to the  $SU(6)_V$  transformations which are broken by the weak gauging.  The discussion in the text suggests that the sections (5) and  (6) are actually absent.}
\label{spsu1}
\end{center}
\end{figure}

\begin{figure}
\begin{center}
\includegraphics[width=4in]{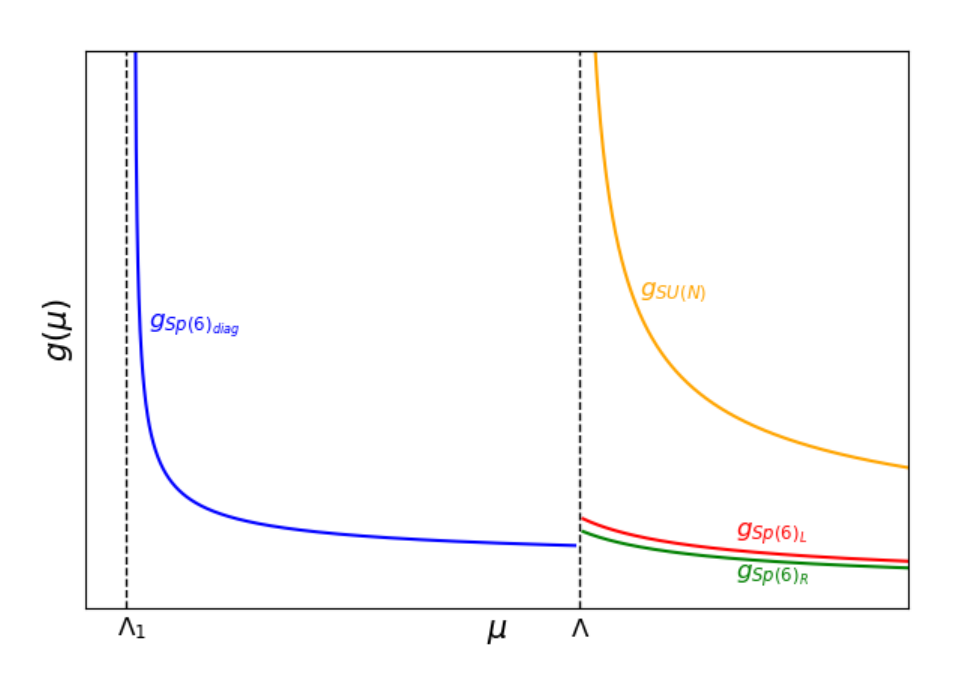}
\caption{\footnotesize  The coupling constant flow of the  $SU(N)- Sp(6)-Sp(6)$  model of Table~\ref{Simplest3}, assuming that $SU(N)$  interactions become strong first at $\Lambda$,  
as the mass scale $\mu$ is lowered.  Below $\Lambda$,  the axial combination of $Sp(6)$ is broken, and the system flows as an $Sp(6)_{\rm diag}$
theory with matter scalars in  $\underline{14}$ of mass $\sim g_{\rm weak} \Lambda$.  }
\label{sunvssp6}
\end{center}
\end{figure}

Symmetries (approximate, weakly gauged, spontaneously broken or unbroken) acting on the fermions in this model are illustrated in  Fig.~\ref{spsu1}. 
  The coupling constant flow looks like Fig.~\ref{sunvssp6}.

\subsection{An $Sp(6)$ getting stronger first}

If one of the $Sp(6)$  factors, e.g.,  $Sp(6)_L$, gets strongly  coupled instead,  when    $SU(N)$ and $Sp(6)_R$ are still weakly  coupled,  it is most likely that
a  condensate 
\be  \brc          \psi  \psi   \ckt,      \label{psipsi}
\ee    
forms which is in the singlet of $Sp(6)_L$.  As it is an antisymmetric product with the metric $\Omega$, as in (\ref{condsp0}) and (\ref{condsp1}),
the VEV (\ref{psipsi}) must also be antisymmetric in $SU(N)$  indices:  it is in the antisymmetric tensor representation
\be     \brc   \psi  \psi    \ckt    \sim \Yvcentermath1   \yng(1,1)\; \label{VEVSp(6)_L}
\ee
of   $SU(N)$.
The VEV (\ref{VEVSp(6)_L}) spontaneously breaks $SU(N)$.  Following \cite{LFLi,Wu:1981eb,Kim:1980ec,Jetzer:1983ij}  we assume that the symmetry breaking pattern is
\begin{equation}
SU(N) \rightarrow Sp(N) \;,  \qquad N=2n \; .   \label{pattern}
\end{equation}
An alternative  symmetry breaking pattern $SU(N) \rightarrow SU(N-2) \times SU(2)$ induced by a scalar VEV  in the two-index antisymmetric representation  has been considered in the literature. In our context, such a breaking pattern seems to be less likely. 
  When the $Sp(6)_L$ interactions become strongly coupled, it will generate the VEV, (\ref{VEVSp(6)_L}),  but also confine. In the decomposition of the $\psi$ field under  $SU(N-2) \times SU(2)$,   $({\underline {N-2}}, {\underline 1}) \oplus  ({\underline {1}}, {\underline 2})$,   only the second component directly participate in the condensate,  but the first component will also be confined, to form a massive $Sp(6)_L$  singlet mesons.  At this point the resulting $SU(N-2)$ gauge symmetry would 
become anomalous, having only massless fermions in $({\underline {N-2}})^*$  from  ${\tilde \psi}$.  
 
With the symmetry breaking,  (\ref{pattern}),  
the  fermionic spectrum  below $\Lambda$  consists solely of  $\tilde{\psi}$ field, as  in Table \ref{repr2}.
\begin{table}[H]
\begin{center}
\begin{tabular}{|c|c|c|}
\hline
$   {}_{\rm field} \backslash {}^{\rm gauge \, group}$ & $Sp(N)$ &  $Sp(6)_R$\\
\hline
&& \\ [-2ex]
$\tilde{\psi}$ & $\underline{N}$  & $\underline{6}$\\
\hline
\end{tabular}
\caption{\footnotesize Representation of the fermionic matter content for energies $\mu \ll  \Lambda_1$.}
\label{repr2}
\end{center}
\end{table}
 Note that this system  is free from both gauge and Witten anomalies.

The one-loop coefficients of the beta functions of $Sp(N)$ and $Sp(6)_R$  are respectively   ($n=N/2$):
\begin{gather}\label{bSp(N)}
b_0[Sp(N)] = - \frac{11n +  5 }{3}    \; ,\\
b_0\left[Sp(6)_R\right] =  - \frac{44-2n}{3}  \; . \label{bSp(6)R}
\end{gather}
In particular, (\ref{bSp(N)}) is negative  for all  $n \ge 2  $, whereas (\ref{bSp(6)R}) is positive for $n > 22$.

Let us assume that the  $Sp(N)$ interactions get strong, at a new mass scale $\Lambda_2 \ll \Lambda_1$.
 Again, a bifermion condensate,
 \begin{equation}
\brc\tilde{\psi}\tilde{\psi}\ckt \propto (\Lambda_2)^3 \neq 0 \; .  \label{it} 
\end{equation}
 is assumed to form. Due to the most attractive channel (MAC) criterion \cite{Raby}, $\tilde{\psi}\tilde{\psi}$ is a singlet of $Sp(N)$, antisymmetric in the   $Sp(N)$ indices.  By statistics, 
 (\ref{it}) must also be antisymmetric in the  $Sp(6)_R$ indices.  The simplest possibility is that it is also an  $Sp(6)_R$  singlet.  $Sp(6)_R$  remains unbroken.   The ${\tilde \psi}$   field is confined and gets a dynamical mass.  
 
\begin{figure}
\begin{center}
\includegraphics[width=4in]{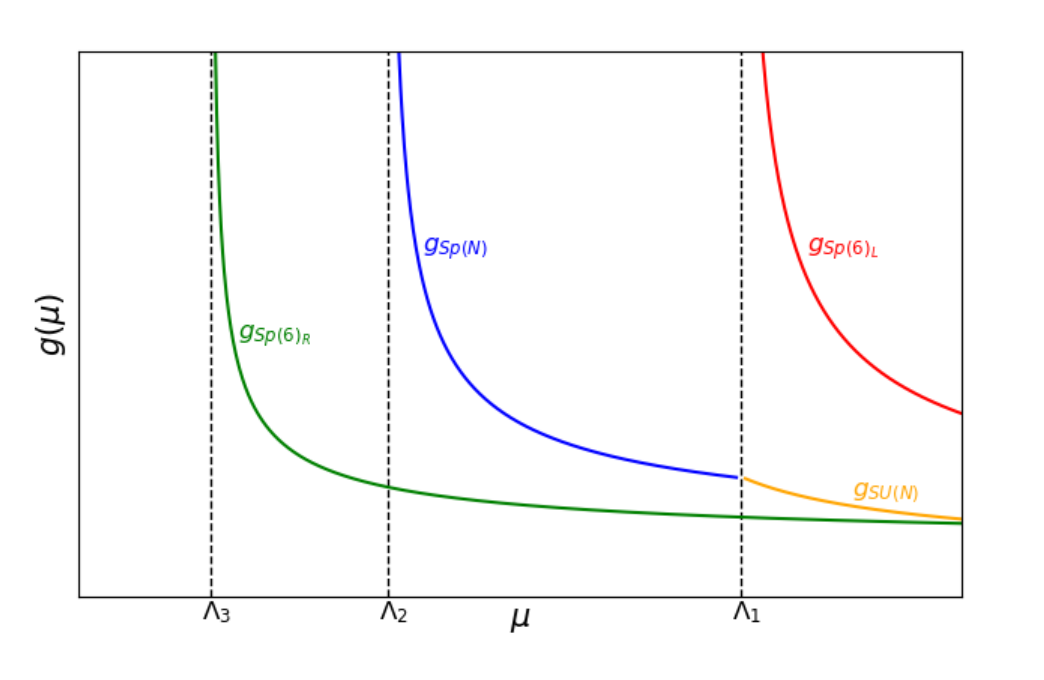}
\caption{ \footnotesize   The coupling constant RG   flow of the  $SU(N)- Sp(6)_L-Sp(6)_R$  model of Table~\ref{Simplest3}, but this time one of the 
 $Sp(6)$ gauge forces, e.g.,  $Sp(6)_L$,  become strong first,  at $\Lambda_1$.  At that scale a bifermion condensate breaks $SU(N)$ to $Sp(N)$,    
which then flows towards the infrared,  getting strongly  coupled at   $\Lambda_2$, whereas   $Sp(6)_R$, unaffected, flows towards the IR  as a pure 
 $Sp(6)_R$  system.  }
 \label{sp6gs}
\end{center}
\end{figure}

 Below   $\Lambda_2$,  the system is described by a pure  $Sp(6)_R$  gauge theory. It will become strongly  coupled and confine at another mass scale, 
 $\Lambda_3 \ll  \Lambda_2 \ll \Lambda_1$.   The RG flow in this case is illustrated  in Fig.~\ref{sp6gs}.

\section{An AF-IF $SU(10)$   model \label{AFIF} }

The final model we consider is  an $SU(10)$ gauge theory  with a single Weyl fermion in the self--adjoint (pseudo--real)
 fifth rank antisymmetric representation, 
 \be     \psi   \sim {\underline {252}}    =  \Yvcentermath1   \yng(1,1,1,1,1) \;,
 \ee 
with the first coefficient of the beta function
\be   b_{0}[SU(10)] = - \frac{ 11\times 10   - 70 }{3}  = -  \frac{40}{3}   <     0 \;.
\ee
At some scale $\Lambda_1$  $SU(10)$  becomes strongly  coupled. The system has been studied recently \cite{BKL1,Yamaguchi},  by making advantage of the 
generalized symmetries and associated new 't Hooft anomaly matching criteria \cite{Seiberg,KapSei,AhaSeiTac,GKSW,GKKS,ShiYon,TanKikMisSak,Komargodski:2017smk,AnbPop1,AnbPop2,Tanizaki,AnberChan,BKLO1}. Here the relevant symmetries are the standard, non-anomalous
${\mathbbm Z}_{70}$  subgroup of $U(1)_{\psi}$,
\be   {\mathbbm Z}_{70}: \qquad    \psi \to      e^{2\pi \im k/70}  \psi \;, \qquad k=1,2,\ldots, 70 
\ee
respected by the instantons,  
and a $1$-form   ${\mathbbm Z}_{5}^C  \subset   {\mathbbm Z}_{10}^C $ center symmetry,   
\be        {\mathbbm Z}_{5}^C:   \qquad     e^{\im \oint  A}  \to         e^{2\pi \im \ell /5}  e^{\im \oint  A} \;,         \qquad \ell  =1,2,\ldots, 5
\ee
acting on the Polyakov (or Wilson) loop  $ e^{i \oint  A} $, but  that does not act on  $\psi$.  By gauging this  $1$-form  ${\mathbbm Z}_{5}^C $ center symmetry one finds that  
(due to the effective fractional instantons)   the  ${\mathbbm Z}_{70}$ symmetry becomes anomalous  (a mixed anomaly), and gets broken as  
\be   {\mathbbm Z}_{70}   \longrightarrow  {\mathbbm Z}_{14}\ .
 \ee
Such a reduction of symmetry means that the vacuum cannot be invariant under  ${\mathbbm Z}_{70}$.   The most natural possibility  \cite{BKL1}  is that
the $\psi$ field condenses
as $SU(10)$  becomes strong,  
\be  \brc \psi\psi \ckt  \ne  0\;.  \label{higher}   \ee
Such a condensate would break ${\mathbbm Z}_{70} $   more strongly,  i.e.,  as     ${\mathbbm Z}_{70}   \to   {\mathbbm Z}_{2}$, than the generalized anomaly suggests, 
 but that is logically consistent.  We are indeed familiar with this situation\footnote{In the standard QCD with $N_f$  massless flavors,  the axial $U(1)_A$  is broken by anomaly and instantons to $ {\mathbbm Z}_{2 N_f}$,  but we know that the effect of the quark condensate $\brc  \bar{\psi} \psi \ckt \ne 0$ is stronger:   
$ {\mathbbm Z}_{2 N_f}$   is broken  by the vacuum to   $ {\mathbbm Z}_{2}$. 
}.

Due to statistics,   $\psi\psi $  condensate cannot be an $SU(10)$ and Lorentz singlet.
  Let us assume that  $\psi\psi$   is in the 
adjoint representation of $SU(10)$.   As
the quadratic Casimir for $\psi$ is quite large,  it is possible that the condensation occurs at mass scale where
\be  \left| \left(C_2({\rm adj}) -  C_2 ( {\underline {252}}) -  C_2 ( {\underline {252}}) \right)   g^2 \right|
=    \frac{35}{2}   g^2      \sim  1\,,
\ee
i.e., at a relatively small  value of $g^2$.

How the   $SU(10)$   
gauge symmetry is broken by the condensate   $\brc \psi\psi \ckt $ in the adjoint representation  is not known.  The discussion of this section
is necessarily of a more speculative nature than that of other sections of this work. 
For definiteness, we assume here that   it breaks the gauge symmetry as 
\be    SU(10) \to   SU (8) \times SU(2) \times U(1) \;, 
\label{breakingSU(10)}
\ee
where $U(1)$ is generated by the charge
\be   Q     =      \left(\begin{array}{cc}      \,{\mathbbm 1}_8 & 0 \\0 & - 4\,{\mathbbm 1}_2  \\ \end{array}\right)\;. 
\ee
It might be thought that all components of the $\psi$ field acquire  Dirac like mass term and decouple, leaving a pure $SU (8) \times SU(2) \times U(1)$ gauge theory below the mass scale $\Lambda_1$.

Actually,  the situation is a little subtler.    Decomposition of the $\psi$ field  as the sum of the irreducible representations of the  
unbroken gauge group $SU (8) \times SU(2) \times U(1)$,  gives  Table~\ref{Simplest333}. 
\begin{table}[h!t]
  \centering 
  \begin{tabular}{|c|c |c|c| }
\hline
$     {}_{\rm fields} \backslash {}^{\rm gauge \, group} $   &  $SU(8) $    &  $ SU(2)$     &   $ U(1) $        \\
 \hline 
 &&&\\[-2ex]
         $ \psi_1 $      &   $ \Yvcentermath1 {\yng(1,1,1,1,1)}$     &   $\underline{1}$    &    $5$          \\[6ex]
                  $ \psi_2 $      &   $ \Yvcentermath1 \yng(1,1,1,1) $     &   $ \Yvcentermath1 \yng(1)$    &    $0$          \\ [6ex]
             $ \psi_3 $      &   $ \Yvcentermath1 \yng(1,1,1) $     &   $\underline{1}$      &    $-5$          \\ [3ex]
 \hline
\end{tabular}  
  \caption{\footnotesize Decomposition of the $\psi$ field under the  
subgroup $SU (8) \times SU(2) \times U(1)$.}
   \label{Simplest333}
\end{table}
It can be seen that  the pair  $(\psi_1, \psi_3)$  can  form an   $SU(8) \times SU(2) \times U(1) $  invariant 
Dirac fermion and therefore  probably get mass dynamically. In contrast, due to statistics,   the  $\psi_2$  field cannot form
an $SU(8) \times SU(2) \times U(1) $ and Lorentz invariant pair $\psi_2 \psi_2$ (such a  composite field would be identically zero; see also (\ref{curious3}) and  (\ref{curious4})  below). This
 means that the fermion  $\psi_2$  remains as a massless degree of freedom,  below  the mass scale,  $\Lambda_1$.

As the fermion pair $\psi_1-\psi_3$ formed a massive meson, neutral of the $U(1)$ charge, and the remaining fermion $\psi_2$ carries no charge with respect to it,
 the $U(1)$ gauge group becomes a free theory,  with its  photon decoupled from the rest of the system. 
 
With only $\psi_2$   as the  matter fermion field,   $SU(8)$ is still asymptotically-free, as
\be  b_8 = -    \frac{11  \times 8   -   20   \times 2   }{3}    = -   16    <    0\;,  \label{1loop1}
\ee
whereas  $SU(2)$   is infrared-free (IF):  
\be  b_2   =   -    \frac{ 11 \times 2   -    70 }{3}      =   +  \frac{68}{3}      >           0  \;.   \label{1loop2}
\ee

\begin{figure}
\begin{center}
\includegraphics[width=5in]{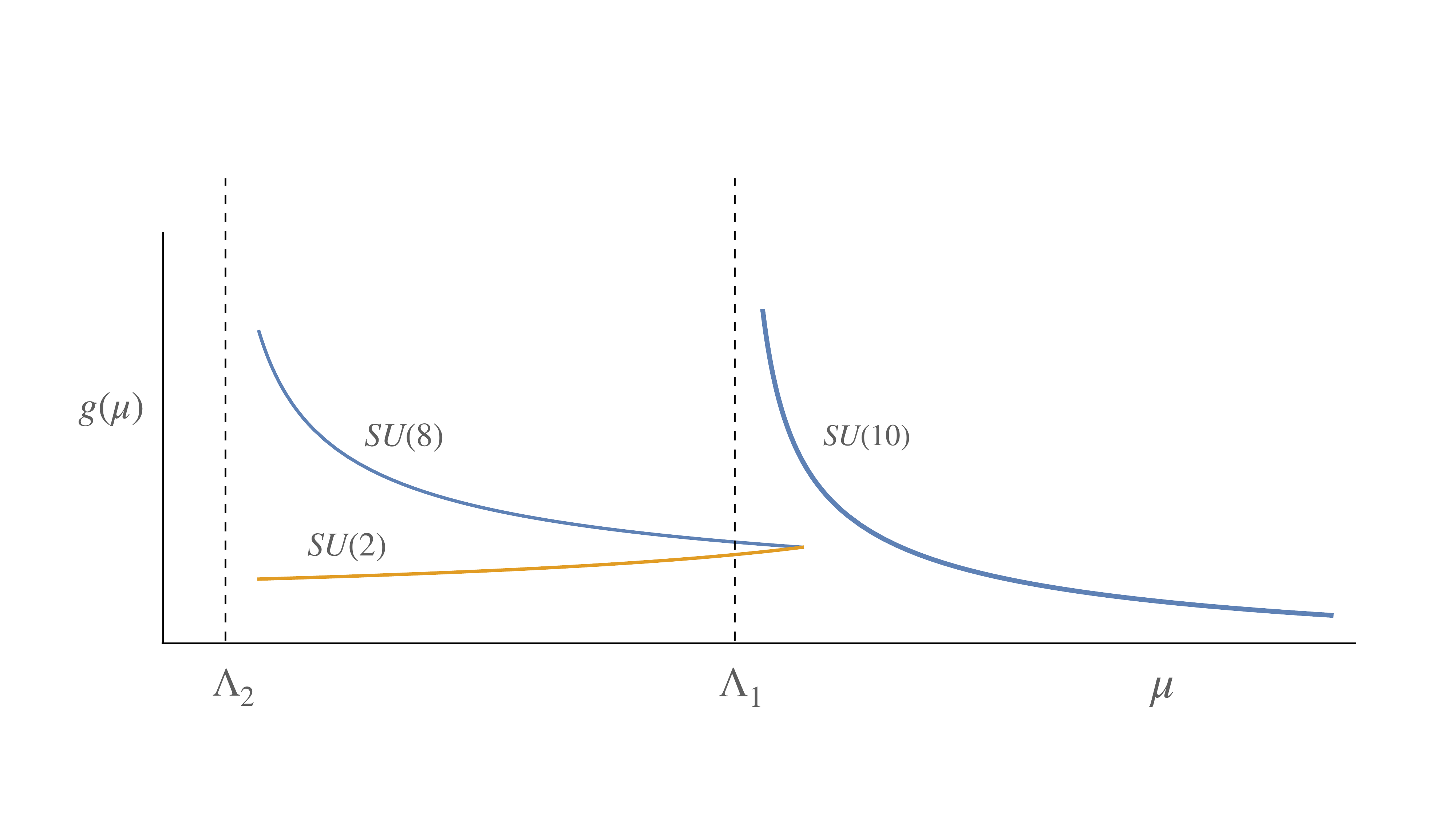}
\caption{ \footnotesize  The coupling constant RG   flow of the  $SU(10)$  model assuming the breaking (\ref{breakingSU(10)}). Due to the large quadratic Casimir the breaking might happen when the coupling constant is still relatively small. The two unbroken subgroups are one AF and the other IF.  }
\label{RGSU10}
\end{center}
\end{figure}

While the $SU(8)$ effective coupling constant  $g_8(\mu)$   grows larger   towards  the infrared,  the   $SU(2)$  coupling gets  weaker.   
The RG flow in this model  is illustrated in Fig.~\ref{RGSU10}.   Even though such a picture is based on the one--loop results,   (\ref{1loop1}), 
 (\ref{1loop2}),    and $g_8$ and $g_2$  evolutions will certainly mix under renormalization beyond the one--loop approximation,  it is probably safe to conclude that the  $SU(8)$  gauge interactions become strongly coupled at some mass scale
\be    (\Lambda_2)^2  =   (\Lambda_1)^2    \,   e^{16\pi^2 /  b_8  \, g_8(\Lambda_1)^2  }    \ll    (\Lambda_1)^2\;.
\ee
It is then natural to assume that 
the $\psi_2$-field condensate will form: 
      \be  \brc \psi_2 \psi_2 \ckt  \ne  0\;.  \label{abovecond}   \ee

      It might be thought that this will give dynamical mass  to the   $\psi_2$  field and wash away any interesting physics  below the mass scale $\Lambda_2$.  Here, again, we find a nontrivial physics. The reason is that the dynamics of the condensate formation (\ref{abovecond})  is now dictated by the strong $SU(8)$  forces,  
in contrast to what happened at the higher mass scale $\Lambda_1$,   (\ref{higher}).     
The $SU(2)$  interactions now play a secondary role.  This means that  the above condensate (\ref{abovecond}) has the $SU(8)$ indices saturated as
\be    \brc \phi^{ij}  \ckt\;, \qquad     \phi^{ij}  \equiv \psi_2^{a_1 a_2 a_3 a_4  i} \psi_2^{a_5 a_6 a_7 a_8  j} \,    \epsilon_{a_1 a_2 \ldots a_8} \,    \;, \label{curious3} 
\ee
i.e., an $SU(8)$ and Lorentz singlet,  but  it  cannot  also be  an $SU(2)$ singlet,  as  
\be   \epsilon_{ij}  \phi^{ij}  \equiv 0\;. \label{curious4} 
\ee
In other words,  (\ref{curious3})  must be an $SU(2)$  triplet.   
That is,   the  $\psi_2 \psi_2 $ condensate  (\ref{abovecond}) breaks  the $SU(2)$ gauge symmetry as 
\be    SU(2) \to U(1)\;.    
\ee
If the $SU(2)$  were a global symmetry,  this would mean the presence of two exactly  massless NG bosons.  But as the $SU(2)$ are weakly-coupled {\it  gauge}   symmetry,  we have dynamical Higgs phenomenon.  Two of the $SU(2)$  gauge bosons   (``W-bosons'') get massive,  with mass of the order of  
\be    \sim      g_2(\Lambda_2)  \Lambda_2 \;,  \ee 
while the third $SU(2)$  gauge boson remains exactly massless:    another  ``photon''.  Note that even if   a $U(1)\subset SU(2)$
gauge symmetry remains unbroken, the fermions $\psi_2$ all get massive, as both $\psi_2^{a_1 \ldots  a_4  i} $,  $i=1,2$,  participate in the condensate and as a result get a dynamical mass.

At the mass scales much below  $g_2 \Lambda_2$,  therefore,  the world is  a free $U(1)\times U(1)$ system with no light  matter coupled to them.  
 The general state describes an arbitrary excitation and number of two types of photons. 

\section{Summary} 
\label{summary}

In this paper we studied the infrared spectra of several simple asymptotically-free chiral gauge theories, with the aim of getting a better grasp on 
the strong gauge dynamics of this type of systems. 
We considered  models with a few nonabelian gauge group factors  and no nonabelian global symmetries.  Depending on the gauge groups involved, on the fermion representations, the relative strength of their coupling constants at some fixed UV mass scale,  varieties of IR effective theories emerge.  
In the deep infrared,  some system reduces to one with a single kind of free massless fermion  (the model of Sec.~\ref{quiver}), or two kinds of free photons (in the system of Sec.~\ref{AFIF}).  In others, the system is gapped. 
At somewhat higher mass scales,  physics we find varies from an $SU(4)$ theory with a self-adjoint antisymmetric tensor fermion  (Sec.~\ref{SUNfirst}),  an $SU(N)$  QCD  with symmetric--tensor  quarks (Sec.~\ref{scenario2}), an $Sp(6)$ theory with massive scalars in ${\underline {14}}$  (Sec.~\ref{SUvsSp}),  a pure ${\cal N}=1$ Supersymmetric $SU(N)$ Yang--Mills theory (Sec.~\ref{quiver}), 
an  $SU(8)\times SU(2)\times U(1)$ theory with an asymptotically-free $SU(8)$ and infrared--free $SU(2)\times U(1)$ gauge interactions  (Sec.~\ref{AFIF}),  and so on.

 Though these results are of some interest,  clearly a more elaborated UV model is needed to produce dynamically  an $SU(3)\times SU(2)\times U(1)$  theory, with  asymptotically--free  ($SU(3)$) and infrared-free ($SU(2)\times U(1)$)  gauge groups,  with three families of quarks and leptons, with some  light effective Higgs scalar bosons doing their required job (of breaking correctly the electroweak gauge symmetry and giving mass to the fermions), as the low--energy effective theory.  In particular, we need a systematic understanding of how the  nontrivial quark  CKM matrix (and the analogous lepton mixing matrix) arises, and how the mechanism generating these,  in general complex,  nonvanishing quark and lepton masses,  leaves  at the same time $\theta_{eff}  \simeq 0$ in the QCD sector.  We hope to come back to discuss these questions in the near future.

\section*{Acknowledgment} 

We thank Andrea Luzio for various discussions on previous related projects.
This work is supported by the INFN special initiative grant, GAST (``Gauge and String Theories'').

\appendix

\section{The gauge-field mixing in the $SU(N)-SU(N+4)$  model   \label{mixing} }

In the dynamical Higgs phenomenon,  an appropriate set of NG bosons are absorbed by the original $SU(N)$ gauge fields, 
making them massive, and leaving orthogonal combinations of $SU(N)\times SU(N+4)$ massless, an exact 
$SU(N)^{\prime} \times SU(4)$ gauge theory.  Let us work out the things explicitly.  

We introduce an effective scalar field representing the composite, 
\be  \phi^i_a    \equiv     \psi^{ij} \eta_{j\,a}  \;,   
\ee
in 
\be  \Yvcentermath1      \yng (1) \; ,  \; {\bar {\yng (1)}}    
\ee
of   $SU(N)\times SU(N+4)$.   It acquires a  VEV of the form, 
\be      \brc \phi \ckt \sim     v  \left(\begin{array}{cc}{\mathbbm 1}_{N\times N} & {\mathbf 0}_{N\times 4}\end{array}\right)\;,\label{VEVphi}
\ee
breaking the symmetries as
\be     SU(N)\times SU(N+4) \times U(1)  \to     
   SU(N)^{\prime}  \times  SU(4) \times U(1)^{\prime}\;. 
\ee
As we noted earlier, in the absence of $SU(N+4)$ gauge interactions ($g_2=0$), there would be $8N+1$ physical massless NG bosons. 
$SU(N)^{\prime}  \times  SU(4) $ is 
an unbroken gauge symmetries, with massless fermions in Table~\ref{LE1}.  
The kinetic term of $\phi$  is 
\be      |\left( \partial_{\mu} - \im g_1  T^A_{\tiny \yng(1)} A^A_{\mu}   -  \im g_2  t^A_{\tiny \bar {\yng(1)}}  B^A_{\mu} \right) \phi |^2\;, \label{kin1}
\ee 
for the part of  $ \phi^j_a$,  $j, a =1,2,\ldots, N$, and    
 \be      |\left( \partial_{\mu} - \im g_1  T^A_{\tiny \yng(1)} A^A_{\mu}   -  \im g_2  t^A_{\tiny {SU(4)}}  B^A_{\mu} \right) \phi |^2\;.  \label{kin2}
\ee 
for the rest  of  $\phi^j_a $,     $j=1,2,\ldots, N, \quad  a=N+1, \ldots, N+4$. 
Under the VEV of the form,  (\ref{VEVphi}),   the kinetic term (\ref{kin1}) gives rise to massive $SU(N)$  gauge bosons for a combination of $SU(N)$ and $SU(N+4)$,  
\be     g G_{\mu}^A \equiv  g_1 A_{\mu}^A -  g_2 B_{\mu}^A\;, \qquad   g\equiv \sqrt{g_1^2+g_2^2}\;, \qquad A =1,2,\ldots, N^2-1\;, 
\ee
of mass, $  g v$. 
The orthogonal combination, 
\be      g C_{\mu}^A \equiv  g_2 A_{\mu}^A +  g_1 B_{\mu}^A\;, \qquad   g \equiv \sqrt{g_1^2+g_2^2}\;,   \qquad A=1,2,\ldots, N^2-1\;,   
\ee
remain massless and gives rise to low--energy $SU(N)^{\prime}$  gauge interactions.  Inverting these,  
\be A_{\mu}^A  \equiv   \frac{1}{g}  ( g_1 G_{\mu}^A +  g_2  C_{\mu}^A)\;, \qquad   B_{\mu}^A  \equiv   \frac{1}{g}  ( -g_2 G_{\mu}^A +  g_1  C_{\mu}^A)\;.
\ee
The $4 \times 4$  generators of $SU(4)\subset SU(N+4)$,     $ B_{\mu}^B$   remain massless and describe the low-energy $SU(4)$
gauge interactions.

There are also   the gauge bosons  $ B_{\mu}^A $  of  $SU(N+4)$,  corresponding to the generators having matrix elements $a=  1,2, \ldots  N, \,\, b=  N+1, N+2,\ldots, N+4$,  or      $a=  N+1, N+2,\ldots, N+4 \,, b= 1,2, \ldots  N $,    which become massive  absorbing the NG bosons. 
Finally  the gauge boson $ B_{\mu}^A $  of  $SU(N+4)$  for the generators  
\be       \left(\begin{array}{cc} 4  {\mathbbm 1}_N & 0 \\ 0 & -N  {\mathbbm 1}_4\end{array}\right)
\ee  
also gets massive eating the last NG boson. 

For simplicity let us write the low-energy effective action for the fermions  taking into account the unbroken $SU(N)^{\prime} \times  SU(4)$  gauge 
interactions only, even though they are in general also coupled to the   $N^2-1+ 8N + 1= N^2+8N $   massive broken gauge bosons, i.e., an anologue of the real--world  ``weak interactions''. 

The kinetic terms of the fermions are
\begin{align} 
D_{\mu}  \psi &=   (\partial_{\mu}  -  \im  g_1  T^A_{\tiny \yng(2)}   A_{\mu}^A ) \psi  \;,  \nonumber \\
D_{\mu}  \eta  &=   (\partial_{\mu}  -  \im  g_1  T^A_{\tiny \bar{ \yng(1)}}   A_{\mu}^A  -    \im g_2   t^A_{\tiny \bar {\yng(1)}} B_{\mu}^A ) \eta \;,  \nonumber \\
D_{\mu}  \chi&=   (\partial_{\mu}  -  \im  g_2  t^A_{\tiny \yng(1,1)}   B_{\mu}^A ) \chi  \;. 
\end{align}
Decomposing the fermions with respect the unbroken symmetry groups,  $SU(N)^{\prime}  \times  SU(4) $ as in Table~\ref{LE1}, 
their kinetic terms become
\begin{align} 
D_{\mu}  \lambda_1 &=   (\partial_{\mu}  -   \im g_2   t^A_{\tiny \bar {\yng(1,1)}} B_{\mu}^A ) \lambda_1=    (\partial_{\mu}  -   \im g^{\prime}   t^A_{\tiny \bar {\yng(1,1)}} C_{\mu}^A ) \lambda_1    \;,   \qquad   t^A \in  \mathfrak{su}(N)^{\prime}\;,    \nonumber \\
D_{\mu}  \lambda_2 &=   (\partial_{\mu}  - \im  g^{\prime}  t^A_{\tiny  \bar {\yng(1)}}  C_{\mu}^A  -  \im g_2  t^B_{\tiny \yng(1)}  B_{\mu}^B )     \lambda_2   \;,  \qquad    t^B\in   \mathfrak{su}(4)\;,  \nonumber \\
D_{\mu}  \chi_1  &=   (\partial_{\mu}  -  \im  g^{\prime}  t^A_{\tiny \yng(1,1)}   C_{\mu}^A ) \chi_1  \;, \nonumber \\
D_{\mu}  \chi_2  &=   (\partial_{\mu}  -  \im  g^{\prime}  t^A_{\tiny \yng(1)}   C_{\mu}^A     - \im g_2   t^B_{\tiny \bar {\yng(1)}}  B_{\mu}^B     ) \chi_2  \;, \nonumber \\
D_{\mu}  \chi_3   &=   (\partial_{\mu}  -  \im  g_2  t^B_{\tiny {\yng(1,1)}}   B_{\mu}^B ) \chi_3  \;, 
\end{align}
where 
\be    g^{\prime} \equiv   \frac{g_1 g_2}{g} \equiv    \frac{g_1 g_2}{\sqrt{g_1^2+g_2^2}} \simeq   g_2 \ll g_1   \;. 
\ee

\section{The gauge-field mixing in the $SU(N)^3$  quiver model  \label{mixing2} }

The effective scalar which acts as the Higgs field is the composite 
 \be  \phi_a^k \sim     \psi^i_a  \kappa^k_i \;.\label{like1} \ee
   It transforms under $SU(N)_2 \times SU(N)_3$  as   (by using a matrix notation), 
   \be    \phi  \to    U_3  \, \phi \,  U_2^{\dagger} \;,   \qquad     U_2 \in   SU(N)_2\;, \quad  U_3 \in   SU(N)_3\,.
   \ee 
   In the kinematic term for   $\phi$,  $\partial_{\mu}  \phi$ transforms as 
   \begin{align}       \partial_{\mu}  \phi    \ \to \  &       \partial_{\mu}     \,  (  U_3  \, \phi \,  U_2^{\dagger}  ) =   \nonumber    \\
      &=        U_3  \,  (\partial_{\mu}  \phi) \,  U_2^{\dagger}  +    U_3 \, {\overleftarrow {\partial_{\mu}}} \,  \phi  \, U_2^{\dagger}   +     U_3  \, \phi \,    \partial_{\mu} U_2^{\dagger}=  \nonumber \\   
      &=     U_3  \,  (\partial_{\mu}  \phi) \,  U_2^{\dagger}  -    U_3 \,  \big( {\partial_{\mu}} \,  U_3^{\dagger} \big)    \,  U_3 \,   
        \phi  \, U_2^{\dagger}   +     U_3  \, \phi \,   U_2^{\dagger}   \,  U_2    \partial_{\mu} U_2^{\dagger}        \label{asin} 
   \end{align}   
In order to have a  locally $SU(N)_2 \times SU(N)_3$ invariant kinematic term,    one must write  the covariant derivative as 
\be     {\cal D}_{\mu} \phi  =   \de_{\mu} \phi     - \im g_3  \,  A_{\mu}^{(3)}   \, \phi   +   \im  g_2  \, \phi   \,  A_{\mu}^{(2)} \;,  \label{covariant} 
\ee
with the gauge field transformations, 
\be        A_{\mu}^{(2)}  \to   U_2  \,(  A_{\mu}^{(2)}   +  \frac{\im}{g_2}  \de_{\mu}   )     U_2^{\dagger}\;, \qquad   
 A_{\mu}^{(3)}  \to   U_3  \,(  A_{\mu}^{(3)}   +  \frac{\im}{g_3}  \de_{\mu}   )     U_3^{\dagger}\;.  \label{like2} 
\ee 
The first term in the kinetic term transforms as in (\ref{asin}).  
The second and third terms in (\ref{covariant}) get transformed as
\begin{align}   & - \im g_3  \,  A_{\mu}^{(3)}   \, \phi  \to  - \im   g_3 \, U_3  \,(  A_{\mu}^{(3)}   +  \frac{\im}{g_3}  \de_{\mu} ) U_3^{\dagger}\, \, U_3  \, \phi \,  U_2^{\dagger} =
\nonumber \\
&=  U_3  \,\left(    - \im g_3  \,  A_{\mu}^{(3)}    +  (\de_{\mu}  U_3^{\dagger} ) \, U_3  \, \right)   \phi \,  U_2^{\dagger}  \;,    \label{term1}
\end{align}
and 
\begin{align}   & \im  g_2  \,   \phi \, A_{\mu}^{(2)}    \to   \im   g_2 \,  U_3  \, \phi \,  U_2^{\dagger}      U_2  \,(  A_{\mu}^{(2)}   +  \frac{\im}{g_2}  \de_{\mu}   )     U_2^{\dagger} =
\nonumber \\
&=  U_3  \,\left(     \im g_2  \,  \phi    \, A_{\mu}^{(2)} U_2^{\dagger}   -  \phi \,  U_2^{\dagger}      U_2  \,     (\de_{\mu}  U_2^{\dagger} )   \right)\;.
 \label{term2}
\end{align}
We note that the second terms in   (\ref{term1}) and  in  (\ref{term2})  cancel the second and the third terms of     (\ref{asin}), respectively.    Thus the kinetic term
of $\phi$,  (\ref{covariant}),     transforms  covariantly as 
\be        {\cal D}_{\mu} \phi  \to     U_3   (  {\cal D}_{\mu} \phi  )    U_2^{\dagger} \;,  
\ee
as desired.  

The  $SU(N)_2 \times SU(N)_3$ gauge fields mix as follows.  
\be    g^{\prime}   B_{\mu}    \equiv       g_2  A_{\mu}^{(2)}   -   g_3   A_{\mu}^{(3)}  \;, \qquad  g^{\prime}  A_{\mu}^{(0)}      \equiv       g_3  A_{\mu}^{(2)}   +  g_2   A_{\mu}^{(3)} \;,  \label{fieldmixing}
\ee
or inverting them,
\be  A_{\mu}^{(2)}       \equiv    \frac{1}{g^{\prime}}      \left(    g_2  B_{\mu}   +    g_3   A_{\mu}^{(0)}  \right)     \;, \qquad    A_{\mu}^{(3)}      \equiv            \frac{1}{g^{\prime}}      \left(    - g_3   B_{\mu}     +    g_2   A_{\mu}^{(0)}  \right)  \;,   \label{invert}  
\ee
where
\be  g^{\prime}   \equiv  \sqrt{g_2^2+ g_3^2}    \;.  \label{SYMcouplingMix}
 \ee
Indeed   in  the Higgs vacuum, (\ref{QC}), 
\be  \brc \phi \ckt \propto  v \,  {\mathbbm 1}\;,    
\ee
we see that  the  $ B_{\mu}  $ field gets mass, 
\be      g^{\prime} v 
\ee
by the Higgs mechanism.   The other field  $A_{\mu}^{(0)}  $  remains massless and describes the unbroken    
$SU(N)_0$ gauge theory.  
The coupling of the  $\lambda$ field with the $SU(N)_0$ gauge field can be found as follows.  
 $\lambda$ field   transforms as  
 \be   \lambda \to    U_2  \lambda   U_3^{\dagger}  \;
 \ee
  (see Table~\ref{Simplest4}).   
 So its kinetic terms is given by   (in the matrix notation) 
 \be     {\cal D}_{\mu}  \lambda =  \de_{\mu}  \lambda   - \im g_2     A_{\mu}^{(2)}   \lambda + \im g_3   \lambda    A_{\mu}^{(3)}  \;. 
 \ee
 and by using the mixing  relation  (\ref{invert}),  we see that the new kinetic term in the theory below   $\Lambda_1$   is given,
 neglecting the coupling to the heavy, broken gauge field  $B_{\mu}$,  simply  by  
\be        {\cal D}_{\mu}  \lambda =    \de_{\mu}  \lambda  -   i g_0   \, [ A_{\mu}^{(0)},  \lambda]   \;,      
 \ee
 where
 \be   g_0 \equiv    \frac{g_2  g_3}{g^{\prime} } \equiv   \frac{g_2  g_3}{\sqrt{g_2^2+ g_3^2}}   \;.  
 \ee
 Note that is is smaller  than either $g_2$ or $g_3$.

\section{A few group-theory formulae for $SU(N)$   \label{Dynkin} }

For convenience of the reader we report here a few group--theory formulae relevant to an $SU(N)$ gauge  theory, 
which were already presented as  an Appendix in \cite{BKL1}.     

 The Dynkin index $T_R$ is defined by
\be   {\rm {tr}} \big( t_R^a t_R^b \big) \equiv  T_R  \,  \delta^{ab}\,,
\ee
where $t_R^a$  are the generators of $SU(N)$ in the representation $R$ . 
Summing over $a=b$, one gets
\be    d(R)  C_2(R) =   T_R \, (N^2-1)\ ,  \qquad     \sum_a t_R^a  t_R^a \equiv C_2(R) {\mathbbm 1}_{d(R)}\ ,
\ee
where $d(R)$ is the dimension of the representation and $C_2(R)$ is the quadratic Casimir. 
For the fundamental representation one has
\be      C_2(F) = \frac{N^2-1}{2N}\ ,  \qquad d(F)=N\ , \qquad   T_{F} =\frac{1}{2}\ ,
\ee
and for the adjoint, 
\be     C_2({\rm adj}) = N\ ,  \qquad d({\rm adj})=N^2-1\ , \qquad   T_{\rm adj} = N\ ;
\ee
these two are quite familiar. 
For a rectangular Young tableau the quadratic Casimir is\footnote{See for example the book of   Barut and Raczka \cite{Barut},  p. 259 (apart from a factor $1/2$ which is included, so that  $C_2(F) = (N^2-1)/2N$ for the fundamental). See also \cite{Walking} for reference.}
 \be   C_2 (  \overbrace{f,  \dots,  f }^k, 0,\dots) =  \frac{k f (N+f) (N-k)}{2 N}\ ,
 \ee 
 where $f$ is the number of the boxes in a row and $k$ the number of rows.

For the order $n$-antisymmetric representation,  $f=1$, $k=n$,      it  is 
\be   C_2(R) = \frac{n (N-n) (N+1)}{2N}\ .
\ee
Taking into account the multiplicity
 \be d(R)=\frac{N(N-1) \cdots (N-n+1)}{n!}=  \frac{N!}{n!  (N-n)!}   \ , 
 \ee
 the Dynkin index of totally antisymmetric single column  representation of height $n$   is given by
 \be    T_R =  \frac{ (N-2)(N-3) \cdots (N-n)}{2 (n-1)!}\ .
\ee

For the special cases  with  $N= 2n$ even,  we have
\be     C_2(R) =   \frac{N(N+1)}{8}\ ,     \qquad   d(R)=      {N \choose  N/2}\ .
\ee
and
\be    2\, T_R =     {N-2 \choose  N/2-1}=   {2n -2 \choose  n-1} \ .\label{thisexp}
\ee

By using this expression    it is easy to see  that  for    $N= 4 \ell$,    $n= 2\ell$,  $2\, T_R$ is a multiple of $ \ell $,  whereas for 
 $N= 4 \ell+2$,   $2\, T_R$  contains $2\ell+1 $ as a divisor. To prove it,  note that the general  combinatoric number
 \be   
   {m \choose  r}=  \frac{m (m-1)\ldots (m-r+1)}{r!} =\frac{m!}{r! (m-r)!}\ , \label{toprove1}
 \ee
 is always an integer.  But  
 \be  {m \choose  r}=  {m \choose  r-1} \cdot \frac{m-r+1}{r}\; 
 \ee
 and both  $ {m \choose  r}$ and $ {m \choose  r-1}$ are integers. 
Therefore  $m-r+1$ is a divisor of  $ {m \choose  r}$.  Applying this for  $N= 4\ell$ one finds that 
\be    2\, T_R =     {4 \ell -2 \choose  2\ell  -1} 
\ee
has a divisor  $2 \ell$ hence $\ell$. 
 For $N=4\ell+2$,  
 \be    2\, T_R =     {4 \ell  \choose  2\ell }    \label{toprove2}
\ee
has a divisor,    $2\ell + 1$.  

For the symmetric representation of rank $2$, $f=2$, $k=1$, so 
\be    C_2(R)= \frac{ (N+2) (N-1)}{N}\ .
\ee
By taking into account the multiplicity,
\be    d(R)=\frac{N(N+1)}{2} \ , 
 \ee
one finds
\be   2  \,T_R=    2  \,   \frac{ d(R)  C_2(R) }{N^2-1}=  N+2\, .
\ee

For the symmetric representation of rank $m$, $f=m$, $k=1$, so 
\be    C_2(R)= \frac{m  (N+m) (N-1)}{
2 N}\ .
\ee
By taking into account the multiplicity,
\be    d(R)=\frac{(N+m-1)!}{m!  \, (N-1)!} \ , 
 \ee
one finds
\be  2\, T_R=      \frac{ d(R)  C_2(R) }{N^2-1}=\frac{(N+m)!}{(N+1)!\, (m-1)!}\ .
\ee
$2\,T_R$  is a multiple of $m$, as can be shown following a similar consideration as  (\ref{toprove1})-(\ref{toprove2}).

\end{document}